\documentclass[aps,pra,twocolumn,groupedaddress,floatfix,nobalancelastpage]{revtex4-1}

\usepackage{graphicx}  
\usepackage{dcolumn}   
\usepackage{bm}        
\usepackage{amssymb}
\usepackage{mathtools}
\usepackage{amsthm}
\usepackage{ragged2e}
\usepackage{verbatim}
\usepackage[colorlinks=true,linkcolor=blue,citecolor=red,plainpages=false,pdfpagelabels]{hyperref}
\usepackage{color,xcolor,colortbl}
\usepackage{multirow}
\usepackage{boxedminipage}
\usepackage{changepage}
\usepackage{bbm}
\usepackage{lipsum}
\usepackage{float}
\usepackage{array}
\usepackage{etoolbox}
\usepackage[caption=false]{subfig}

\robustify{\subref}

\newcommand{\bra}[1]{\langle #1|}
\newcommand{\ket}[1]{|#1\rangle}

\newcommand{\Tr}{\text{Tr}}

\newcommand{\mc}[1]{\mathcal{#1}}
\renewcommand{\t}{{\scriptscriptstyle\mathsf{T}}}
\DeclarePairedDelimiter{\ceil}{\lceil}{\rceil}

\newcommand{\e}{\text{e}}

\def\({\left(}
\def\){\right)}
\def\[{\left[}
\def\]{\right]}

\begin{document}

\widetext


\title{Robust quantum network architectures and topologies for entanglement distribution}


\author{Siddhartha Das}\email{sdas21@lsu.edu} \affiliation{Hearne Institute for Theoretical Physics, Department of Physics and Astronomy, Louisiana State University, Baton Rouge, Louisiana, 70803, USA}
\author{Sumeet Khatri}\email{skhatr5@lsu.edu} \affiliation{Hearne Institute for Theoretical Physics, Department of Physics and Astronomy, Louisiana State University, Baton Rouge, Louisiana, 70803, USA}
\author{Jonathan P. Dowling} \affiliation{Hearne Institute for Theoretical Physics, Department of Physics and Astronomy, Louisiana State University, Baton Rouge, Louisiana, 70803, USA}      

\date{\today}

\begin{abstract}
	
	
	Entanglement distribution is a prerequisite for several important quantum information processing and computing tasks, such as quantum teleportation, quantum key distribution, and distributed quantum computing. In this work, we focus on two-dimensional quantum networks based on optical quantum technologies using dual-rail photonic qubits for the building of a fail-safe quantum internet. We lay out a quantum network architecture for entanglement distribution between distant parties using a Bravais lattice topology, with the technological constraint that quantum repeaters equipped with quantum memories are not easily accessible. We provide a robust protocol for simultaneous entanglement distribution between two distant groups of parties on this network. We also discuss a memory-based quantum network architecture that can be implemented on networks with an arbitrary topology. We examine networks with bow-tie lattice and Archimedean lattice topologies and use percolation theory to quantify the robustness of the networks. In particular, we provide figures of merit on the loss parameter of the optical medium that depend only on the topology of the network and quantify the robustness of the network against intermittent photon loss and intermittent failure of nodes. These figures of merit can be used to compare the robustness of different network topologies in order to determine the best topology in a given real-world scenario, which is critical in the realization of the quantum internet.
	
	
	
	


\end{abstract}

\maketitle

\section{Introduction}
	
	The building of quantum networks is an essential ingredient in the realization of the quantum internet \cite{Kim08}, an interconnected network of quantum networks in which all parties can perform quantum information processing and quantum computing tasks. Execution of many of these tasks is contingent on the reliable distribution of entanglement among the members of the network, such as quantum teleportation \cite{BBC+93,BFK00}, quantum key distribution for secure communications \cite{BB84,GRG+02,SBPC+09}, distributed quantum computation \cite{CEHM99}, Bell inequality tests \cite{AGR82,HBD+15,RBG17}, quantum clock synchronization \cite{JADW00,UD02,ITDB17}, and quantum secret sharing \cite{HBB99}. 
	

	Quantum repeaters \cite{BDC98,DBC99} are essential to overcome the decoherence of particles caused by the environment for reliable entanglement distribution between two parties that are separated by a distance longer than the decoherence length of the communication channel. Much like classical repeaters, which are placed at intermediate points along the channel in order to amplify the signal being transmitted, quantum repeaters employ entanglement swapping \cite{BBC+93,ZZH93} and optionally entanglement purification \cite{BBP96,DAR96,BDD96} at intermediate points along the quantum channel in order to increase the likelihood of establishing entanglement and to increase the fidelity of the entanglement. 
	
	Entanglement purification protocols require the use of quantum memories, which are not widely available with current technologies, and will be expensive in the near-term once they are widely available. It is therefore of interest to build networks and to devise protocols that use quantum repeaters without quantum memories, which we will refer to as memory-free quantum repeaters (see Refs. \cite{KKL15,MEL17} for examples of memory-free quantum repeaters).  
	
	Using optical fiber, a common medium used to transmit quantum signals, with single photons as the qubits, photon loss is the most dominant source of noise, and entanglement purification is not necessary unless other general sources of errors are also considered. (Quantum memories may still be required, however, to store the qubits for later processing.) One drawback to using optical fibers, however, is the known exponential decay with distance of both the probability of successfully transmitting a photon over the fiber and the rate of entanglement generation between the two ends of the fiber \cite{TGW15,PLOB17}.
	
	In order to mitigate this exponential decay of the entanglement-generation rate for point-to-point links, suppose that the two parties that would like to share long-distance entanglement are groups consisting of several spatially-distributed members. The two groups could be, for example, two companies, and the members of each group could be the branches of the companies. In Section \ref{sec-memory_free}, we provide an example of a two-dimensional (2D) lattice-based network topology, in which the branches of each company are at the two opposite ends of the network. The network consists of source stations producing entangled dual-rail photonic qubits transmitted over optical fiber and measurement stations performing entanglement swapping. In Section \ref{subsec:bravais_lattice}, we exhibit a simultaneous entanglement distribution protocol on this network to show that if we care only about creating entanglement between the two groups (and not about creating entanglement between particular members), then the average yield of entangled pairs of photons shared between the two groups is greater than what can be achieved with a single channel between the two groups. We consider only pure loss as the source of noise and do not require the use of quantum memories even to store the qubits temporarily.
	
	
	Another major concern in any network, including the internet, is its vulnerability to failures of, or attacks on, some of the nodes in the network. The internet is extremely robust and can function even when a significant fraction of the nodes fails. When realized, the quantum internet should possess a similar robustness to attacks or to failures of its nodes. How should our networks be constructed so that the quantum internet is fail-safe? In Section \ref{sec-gen_arc}, we address this concern using techniques from bond and site percolation theory by defining figures of merit for networks based on the critical bond and site percolation probabilities of their corresponding graph. We calculate these figures of merit for network topologies obeying certain symmetries and use them to compare the robustness of the networks against photon loss and failures of the nodes. Our figures of merit allow for a direct comparison of different networks for the purpose of long-distance entanglement distribution that is based solely on the topology of the network.
	
	The network architecture presented in Section \ref{sec-memory_free} can be used to perform more than just the protocol presented in Section \ref{subsec:bravais_lattice}. As we describe in Section \ref{sec-discussion}, the network architectures presented in both Section \ref{sec-memory_free} and Section \ref{sec-gen_arc} can be generalized to perform entanglement routing as well as to execute entanglement distribution protocols involving matter qubits.

	

\section{Quantum network with memory-free quantum repeaters}\label{sec-memory_free}

	As a precursor to a full-fledged quantum network, let us consider the following scenario. Suppose two trusted parties, company $X$ and company $Y$, each with one head office that are at distant locations, would like to communicate securely with each other using current (or near-term) quantum technologies. They could execute, for example, an entanglement-based quantum key distribution protocol, which requires the companies to first share entanglement. Suppose that the distribution of entanglement between the companies is to be carried out using single photons over optical fiber. We model the loss over the optical fiber using the pure-loss bosonic channel, also called the attenuator, to be defined below \eqref{def:pure-loss-bc}. Given the fact that quantum memories are currently not widely accessible, due to technological limitations, we treat quantum memories to be highly expensive. This limits the companies to communicate securely by making use of memory-free quantum repeaters, i.e., entanglement swapping stations along the fiber connecting the two companies, to share maximally entangled photon pairs.
	
	Now, there is a limit to the optimal yield of entangled photons between the two ends of an optical fiber if only the head offices are connected by a single fiber. The probability that an entangled pair of photons is shared between the two ends of the fiber of length $L$ is $\eta=\e^{-\alpha L}$ \cite{GPLS09,TGW15,PLOB17,WTB17}, where $\alpha>0$ is a parameter that depends on the property of the fiber. The maximum rate of entanglement generation between the two ends, without any repeaters, is $-\log_2(1-\eta)$ entangled pairs per optical mode \cite{PLOB17,WTB17}. This direct generation of entangled photons over an optical fiber performs no worse than when entanglement-generating sources and memory-free quantum repeaters are placed between the two ends.
	
	However, suppose that each company has, in addition to a head office, several branches that are allowed to perform tasks on behalf of the head office, so that entanglement between any two branches of the companies is sufficient for communication. Can the yield of entangled photon pairs be increased in this case? We show in this section that the yield of entangled photon pairs can be increased.


\subsection{Network architecture and entanglement distribution protocol}\label{subsec:bravais_lattice}
	
\begin{figure}
		\centering
		\includegraphics[scale=0.7]{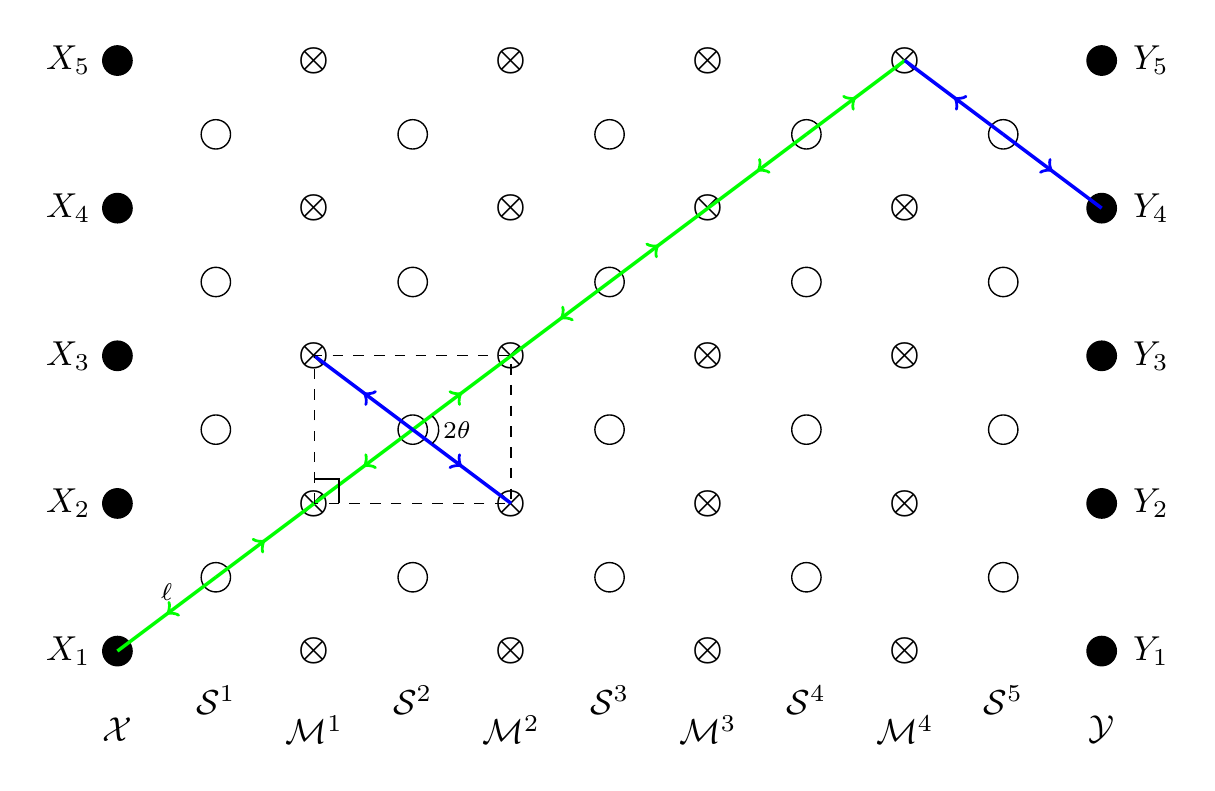}
		\caption{A two-dimensional network with the topology of a Bravais lattice containing $N=5$ branches of company $X$ and company $Y$ as well as $M=4$ columns of measurement terminals. Indicated in blue is the firing of a pair of entangled photons in the anti-diagonal direction, while the green line indicates the simultaneous firing of photons in the diagonal direction, with the final source $S_4^5$ firing in the anti-diagonal direction, in order to create entanglement between $X_1$ and $Y_4$.}
		\label{fig-grid}
	\end{figure}

	Consider a network that is a 2D grid with a rectangular centered Bravais lattice structure, see Fig. \ref{fig-grid}, consisting of an equal number of branches of company $X$ and company $Y$, with source stations for entanglement generation and measurement terminals for entanglement swapping. There are $N$ branches of company $X$ ($Y$), and they are labeled by $X_i$ ($Y_i$), $i\in\{1,2,\dotsc,N\}$, and represented by $\bullet$. The measurement terminals are represented by $\otimes$ and labeled by $M^i_j$, $i\in\{1,2,\dotsc, M\}$, $M$ being the number of measurement terminals in each row of the grid, and $j\in\{1,2,\dotsc, N\}$ denoting the rows of the grid. Similarly, the source stations are represented by $\circ$ are labeled by $S^i_j$, $i\in\{1,2,\dotsc, M+1\}$ denoting the source stations in each row of the grid and $j\in\{1,2,\dotsc, N-1\}$ denoting the rows of source stations. We define the following sets:
	\begin{align}
		\mathcal{X}&=\{X_1,X_2,\dotsc,X_N\},\\
		\mathcal{Y}&=\{Y_1,Y_2,\dotsc,Y_N\},\\ 
		\mathcal{S}^i&=\{S_1^i,S_2^i\dotsc, S_{N-1}^i\}~\forall ~i\in\{1,2,\dotsc, M+1\},\\
		\mathcal{M}^i&=\{M_1^i,M_2^i\dotsc, M_N^i\}~\forall~i\in\{1,2,\dotsc, M\}.
	\end{align} 
	We let $\mathcal{X}\equiv\mathcal{M}^0$ and $\mathcal{Y}\equiv \mathcal{M}^{M+1}$. The area covered by the network is $LH$, where $L=2\ell(M+1)\cos\theta$ is the total horizontal length of the network and $H=2\ell(N-1)\sin\theta$ is the total vertical length of the network.
	
	Since quantum memories are not widely accessible and are expensive, we suppose that only the branches $X_i$ and $Y_i$ have quantum memories, while the measurement terminals contain memory-free quantum repeaters.
	
	We use a dual-rail scheme based on single photons to encode the qubits and optical fibers to transmit the photons among the nodes in the network. Let $A_1,A_2$ be two orthogonal optical modes. The dual-rail encoding of a qubit in these two modes is defined by letting the states $\ket{1,0}_{A_1A_2}$ and $\ket{0,1}_{A_1A_2}$, i.e., occupation of either mode by a single photon, represent the computational basis of the qubit system. Specifically, we can let $A_1$ and $A_2$ be two polarization modes of light, so that the computational basis is given by
	\begin{equation}\label{eq-qubit_system_dualrail}
		\ket{H}_A\coloneqq\ket{1,0}_{A_1A_2},\quad \ket{V}_A\coloneqq\ket{0,1}_{A_1A_2},
	\end{equation}
	where $A$ denotes the qubit system and $H$ and $V$ refer to horizontal and vertical polarization, respectively. Though we consider for concreteness throughout this paper polarization-based dual-rail photons defined in this way, our results will hold for other dual-rail encodings, such as when $A_1$ and $A_2$ are frequency-offset modes \cite{RP10}.
	
	We restrict inputs to the optical fiber to the one-photon subspace spanned by $\ket{H}_A$ and $\ket{V}_A$. Then, any pure state $\ket{\psi}_{A_1A_2}$ of the qubit system can be written as
	\begin{align}
		\ket{\psi}_{A_1A_2}&=\alpha\ket{1,0}_{A_1A_2}+\beta\ket{0,1}_{A_1A_2}\\
		&=\alpha\ket{H}_A+\beta\ket{V}_A,
	\end{align}
	such that $|\alpha|^2+|\beta|^2=1$.

\begin{figure}
		\centering
			\subfloat[]{\label{subfig-source_diag_antidiag}\centering\includegraphics[scale=0.7,width=0.13\textwidth]{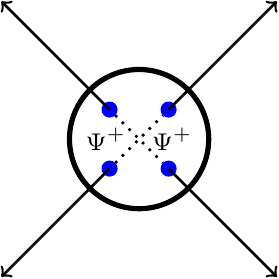}}~~
			\subfloat[]{\label{subfig-source_top_bottom}\centering\includegraphics[scale=0.7,width=0.13\textwidth]{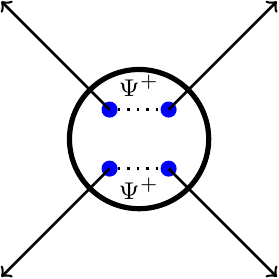}}~~
			\subfloat[]{\label{subfig-source_left_right}\centering\includegraphics[scale=0.7,width=0.13\textwidth]{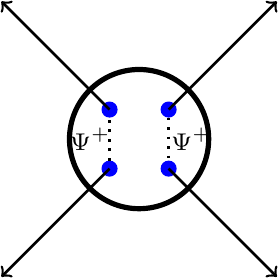}}
		\caption{All source stations in the network create two pairs of photons, with each pair in the Bell state $\Psi^+$. The entanglement is created either between: \subref{subfig-source_diag_antidiag} diametrically opposite photons; \subref{subfig-source_top_bottom} top and bottom photons; or \subref{subfig-source_left_right} the left and right photons.}
		\label{fig-sources}
	\end{figure}

	Each source station generates two pairs of the same Bell state $\Psi^+\coloneqq \ket{\Psi^+}\bra{\Psi^+}$, where $\ket{\Psi^+}=\frac{1}{\sqrt{2}}(\ket{H,V}+\ket{V,H})$. As illustrated in Fig. \ref{fig-sources}, the source station creates the entanglement between either the diametrically opposite photons (\ref{subfig-source_diag_antidiag}), the photons at the top and bottom (\ref{subfig-source_top_bottom}), or the photons at the left and right (\ref{subfig-source_left_right}). Both photons of a pair are fired in opposite directions to the nearest measurement terminals. In Fig. \ref{fig-grid}, photons are shown to be fired as per Fig. (\ref{subfig-source_diag_antidiag}).
	
	
\begin{figure}
		\centering
			\subfloat[]{\label{subfig-diag_antidiag}\centering\includegraphics[scale=0.45,width=0.13\textwidth]{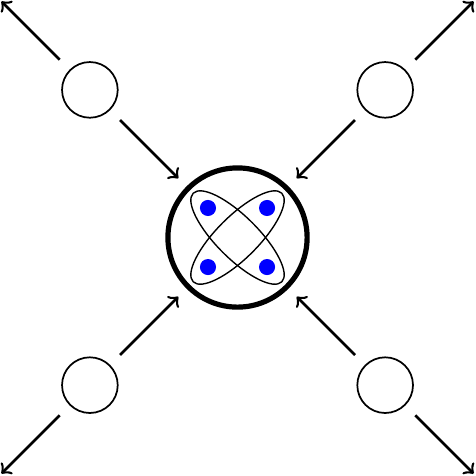}}~~
			\subfloat[]{\label{subfig-top_bottom}\centering\includegraphics[scale=0.45,width=0.13\textwidth]{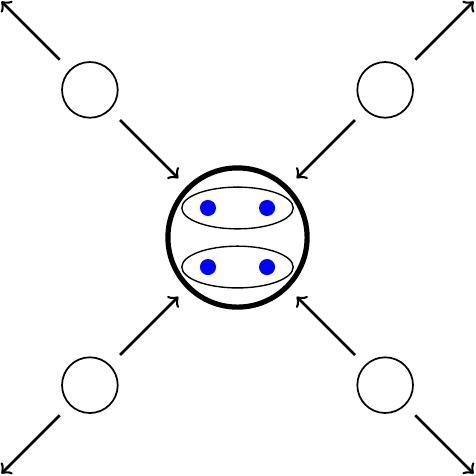}}~~
			\subfloat[]{\label{subfig-left_right}\centering\includegraphics[scale=0.45,width=0.13\textwidth]{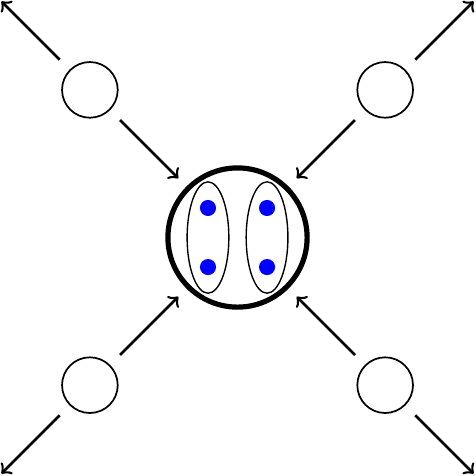}}
		\caption{Pairs of photons (indicated in blue) arriving from the source stations (small outer circles) at a measurement terminal (large central circle). Encircled photon pairs are measured. The measurement terminal measures either: \subref{subfig-diag_antidiag} pairs of diametrically opposite photons; \subref{subfig-top_bottom} pairs of top and bottom photons; or \subref{subfig-left_right} pairs of left and right photons.}
		\label{fig-meas}
	\end{figure}

	Each measurement terminal contains two memory-free repeaters. As illustrated in Fig. \ref{fig-meas}, each measurement terminal has the ability to perform two-photon measurements either on the diametrically opposite photon pairs, on the photons pairs at the top and bottom of the measurement terminal, or on the photons pairs at the left and right of the measurement terminal. Measurement terminals at the edges of the graph receive only two photons and can therefore measure only those two photons. Each repeater performs a Bell measurement with success probability $\gamma$ on the photon pairs it receives. This functionality of the measurement terminals makes them essentially equivalent to quantum relays \cite{JPF02,DMT+04,CGD05,SSR+11}.
	
	Our protocol for distributing entanglement between company $X$ and company $Y$ is the following. At each time step:
	\begin{enumerate}
		\item All source stations fire pairs of entangled photons in the state $\Psi^+$ as per Fig. \ref{subfig-source_diag_antidiag}.
		\item Upon receiving the photons, the measurement terminals immediately perform a Bell measurement on the pairs of photons in the orientation of Fig. \ref{subfig-diag_antidiag}. They all globally announce the results along with a label uniquely identifying the measurement terminal.
		\item The $X$ branches, using the announced measurement results, perform appropriate local operations to recover the state $\Psi^+$.
	\end{enumerate}
	
	Note that the $X$ and $Y$ branches know with which branch of the other company they are entangled by using the announcements from the measurement terminals. Also, for simplicity of the analysis below, in this protocol we suppose that all the entanglement-generating source stations and the measurement terminals are secured from any infiltration by unauthorized parties. A more sophisticated analysis, assuming that the measurement terminals and/or the source stations are untrusted, can be carried out and can potentially allow for the network to be used for device-independent and measurement-device-independent quantum key distribution \cite{MY98,ABG+07,LCQ12,BP12}.
	
	The following subsections are devoted to analyzing this protocol, proving that the branches of company $X$ and company $Y$ will indeed share entanglement after each time step, and determining the average number of entangled photon pairs shared by company $X$ and $Y$ after each time step. 

\subsection{Entanglement transmission from the source stations}\label{sec-ent_dist}
	
	Any physical transformation in quantum mechanics is described by a completely positive and trace-preserving map, also referred to as a quantum channel. We model the transmission of a photon through the optical fiber as a pure-loss bosonic channel, also called attenuator or beam-splitter, which induces the following transformation on the input's annihilation operators $\hat{a}_i$ and the associated environment's annihilation operators $\hat{e}_i$:
	\begin{align}
		\hat{a}_i &\mapsto \sqrt{\eta} \hat{a}_i+\sqrt{1-\eta}\hat{e}_i,\nonumber\\
		\hat{e}_i &\mapsto \sqrt{1-\eta}\hat{a}_i + \sqrt{\eta} \hat{e}_i.
	\end{align}
	This input-output Heisenberg-picture relation is equivalent to conjugation of the annihilation operators by the unitary operator 
	\begin{equation}
		U_i=\exp\left[\cos^{-1}\(\sqrt{\eta}\)\(\hat{a}^\dag_i\hat{e}_i-\hat{a}_i\hat{e}^\dag_i\)\right].
	\end{equation}
	Using this fact, the pure-loss bosonic channel $\mc{E}_{A_1 A_2\to B_1 B_2}$ can be defined as the following quantum channel \cite{BH14}:
	\begin{align} \label{def:pure-loss-bc}
		&\mc{E}_{A_1 A_2\to B_1 B_2}(\rho_{A_1A_2}) \nonumber\\
		&\,\coloneqq \Tr_{E_1 E_2}\left\{U_1^\dag\otimes U_2^\dag \(\rho_{A_1A_2}\otimes \ket{0,0}\bra{0,0}_{E_1 E_2}\)U_1\otimes U_2\right\},
\end{align}
	where for $i\in\{1,2\}$, $A_i$ is the input mode, $B_i$ and $E_i$ are the modes of the output and environment associated with $A_i$. For any state $\rho_{A_1A_2}\equiv\rho_A$ in the qubit subspace spanned by the states $\ket{1,0}_{A_1A_2}=\ket{H}_A$ and $\ket{0,1}_{A_1A_2}=\ket{V}_A$ as defined in \eqref{eq-qubit_system_dualrail}, it holds that \cite{BH14}
	\begin{align}
		\mathcal{E}_{A_1A_2\to B_1B_2}(\rho_{A_1A_2})&=\eta\rho_{B_1B_2}+(1-\eta)\ket{0,0}\bra{0,0}_{B_1B_2} \label{eq:attenuator-channel} \\
		&=\eta \rho_B+(1-\eta)\ket{e}\bra{e}_B,
	\end{align}
	where $\ket{e}_B\coloneqq\ket{0,0}_{B_1B_2}$. The action of the channel $\mc{E}$ on the qubit system $A$ is such that it outputs the exact input state with probability $\eta$ or replaces it with the vacuum state of the two modes $A_1,A_2$ with probability $1-\eta$. Note that the vacuum state is orthogonal to any state of the qubit system since the qubit system is defined on the single-photon subspace of the two modes as per \eqref{eq-qubit_system_dualrail}. The action of $\mathcal{E}$ on the qubit system $A$ can thus be identified with that of the erasure channel with erasure parameter $1-\eta$ and erasure state $\ket{e}\bra{e}$ \cite{GBP97}.
	
	For $X_A\in\{\ket{H}\bra{H}_A,\ket{H}\bra{V}_A,\ket{V}\bra{H}_A,\ket{V}\bra{V}_A\}$, we have
	\begin{equation}\label{eq-erasure_basis}
		\mathcal{E}_{A\to B}(X_A)=\eta X_B+(1-\eta)\Tr(X_B)\ket{e}\bra{e}_B.
	\end{equation}
	
	The four maximally entangled states, also called Bell states, in the space of two qubits are
	\begin{align}
		\ket{\Psi^{\pm}}_{A\bar{A}}&\coloneqq\frac{1}{\sqrt{2}}(\ket{H,V}_{A\bar{A}}\pm\ket{V,H}_{A\bar{A}}),\\
		\ket{\Phi^{\pm}}_{A\bar{A}}&\coloneqq\frac{1}{\sqrt{2}}(\ket{H,H}_{A\bar{A}}\pm\ket{V,V}_{A\bar{A}}),
	\end{align}
	where $\bar{A}$ is another qubit system. Using \eqref{eq-erasure_basis}, the action of the attenuator on each of the systems $A$ and $\bar{A}$ is
	\begin{widetext}
	\begin{align}
		&\mc{E}_{A\to B}\otimes \mc{E}_{\bar{A}\to\bar{B}}(\ket{\Psi^+}\bra{\Psi^+}_{A\bar{A}})\nonumber\\
		&\qquad\qquad =\eta^2\ket{\Psi^+}\bra{\Psi^+}_{B\bar{B}}+\eta(1-\eta)\(\frac{1}{2}\mathbbm{1}_B\otimes\ket{e}\bra{e}_{\bar{B}}+\ket{e}\bra{e}_B\otimes\frac{1}{2}\mathbbm{1}_{\bar{B}}\)+(1-\eta)^2\ket{e}\bra{e}_B\otimes\ket{e}\bra{e}_{\bar{B}},
	\end{align}
	\end{widetext}
	where
	\begin{equation}
		\frac{1}{2}\mathbbm{1}_A=\frac{1}{2}\(\ket{H}\bra{H}_A+\ket{V}\bra{V}_A\)
	\end{equation}
	is the maximally mixed state. Observe that with probability $\eta^2$ $B$ and $\bar{B}$ are maximally entangled, with probability $\eta(1-\eta)$ one of the photons is lost and the other is in a maximally mixed state, and with probability $(1-\eta)^2$ both photons are lost. 
	
	We let
	\begin{align}
		\tau^\eta_{B\bar{B}}&\coloneqq\mathcal{E}_{A\to B}\otimes\mathcal{E}_{\bar{A}\to\bar{B}}(\ket{\Psi^+}\bra{\Psi^+}_{A\bar{A}})\nonumber\\
		&=\eta^2\ket{\Psi^+}\bra{\Psi^+}_{B\bar{B}}+(1-\eta^2)\Psi_{B\bar{B}}^{\perp},\label{eq-Bell_erasure}
	\end{align}
	where
	\begin{equation}\label{eq-Bell_states1}
		\begin{aligned}
		&\Psi_{B\bar{B}}^{\perp}\coloneqq \frac{\eta}{1+\eta}\left(\frac{1}{2}\mathbbm{1}_B\otimes\ket{e}\bra{e}_{\bar{B}}\right.\nonumber\\
		&+\left.\ket{e}\bra{e}_B\otimes\frac{1}{2}\mathbbm{1}_{\bar{B}}\right)+\frac{1-\eta}{1+\eta}\ket{e}\bra{e}_B\otimes\ket{e}\bra{e}_{\bar{B}}.
		\end{aligned}
	\end{equation}
	Note that $\Psi_{B\bar{B}}^{\perp}$ is orthogonal to $\Psi^+$, i.e., $\bra{\Psi^+}\Psi^{\perp}\ket{\Psi^+}=0$. Along with classical communication between $B$ and $\bar{B}$ on whether the photons arrived, the state \eqref{eq-Bell_erasure} is consistent with the action of the erasure channel with erasure parameter $1-\eta^2$ and erasure state $\Psi^{\perp}_{B\bar{B}}$.

\subsection{Entanglement swapping at the measurement terminals}\label{sec:ent_swap}

	All four Bell states in \eqref{eq-Bell_states1} can be written as
	\begin{align}\label{eq-Bell_states}
		\ket{\Phi_{a,b}}&\coloneqq (\sigma_x^a\sigma_z^b\otimes\mathbbm{1})\ket{\Psi^+},
	\end{align}
	where $\sigma_x=\ket{H}\bra{V}+\ket{V}\bra{H}$ and $\sigma_z=\ket{H}\bra{H}-\ket{V}\bra{V}$ are the Pauli-$x$ and Pauli-$z$ operators, and $a,b\in\{0,1\}$. Now, suppose an entanglement source produces a pair of photons in the Bell state $\ket{\Phi_{a_1,b_1}}_{B_1\bar{B}_1}$ and sends one of the photons to $B_1$ and the other photon to $\bar{B}_1$. Similarly, another source distributes a photon pair in the Bell state $\ket{\Phi_{a_2,b_2}}_{B_2\bar{B}_2}$ to $B_2$ and $\bar{B}_2$. If a Bell measurement is performed on the photons $\bar{B}_1$ and $\bar{B}_2$, then it is straightforward to show (see Appendix \ref{appendix-bell_meas_calc}) that each outcome $(a_3,b_3)\in\{0,1\}^2$ occurs with probability $\frac{1}{4}$ and that the corresponding post-measurement state is $\ket{\Phi_{a_1\oplus a_2\oplus a_3,b_1\oplus b_2\oplus b_3}}$, where $\oplus$ denotes addition modulo two.
	
	Using this result, we can determine the state shared by a branch of company $X$ and a branch of company $Y$ along any path in the network. Specifically, let us determine the state shared by $X_1$ and $Y_4$ along the path shown in Fig. \ref{fig-grid} after a single run of the protocol. After the photons arrive at the measurement terminals, the total joint state is
	\begin{equation}\label{eq-joint_state_pre_meas}
		\tau_{X_1M_{2,1}^1}^\eta\otimes\tau_{M_{2,2}^1M_{3,1}^2}^\eta\otimes\tau_{M_{3,2}^2M_{4,1}^3}^\eta\otimes\tau_{M_{4,2}^3M_{5,1}^4}^\eta\otimes\tau_{M_{5,2}^4Y_4}^\eta,
	\end{equation}
	where the notation $M_{j,1}^i$ refers to the photon at the measurement terminal $M_j^i$ arriving from source station $S_{j-1}^i$ and $M_{j,2}^i$ refers to the photon at the measurement terminal $M_j^i$ arriving from source station $S_{j+1}^{i+1}$. Bell measurements are then performed on the pairs $(M_{2,1}^1,M_{2,2}^1)$, $(M_{3,1}^2,M_{3,2}^2)$, $(M_{4,1}^3,M_{4,2}^3)$, $(M_{5,1}^4,M_{5,2}^4)$ at the corresponding measurement terminals. Since each state $\tau^\eta$ contains a term supported on the zero-photon subspace, when measuring the joint state \eqref{eq-joint_state_pre_meas} in the Bell basis on the single-photon subspace spanned by $\{\ket{H},\ket{V}\}$ the only term that will have a non-vanishing contribution to the measurement outcome probabilities and the post-measurement states is the term
	\begin{align}
		&(\eta^2)^5\Psi^+_{X_1M_{2,1}^1}\otimes\Psi^+_{M_{2,2}^1M_{3,1}^2}\otimes\Psi^+_{M_{3,2}^2M_{4,1}^3}\nonumber\\
		&\qquad\qquad\qquad\quad\otimes\Psi^+_{M_{4,2}^3M_{5,1}^4}\otimes\Psi^+_{M_{5,2}^4Y_4}.
	\end{align}
	Thus, after a Bell measurement at each measurement terminal, the post-measurement state corresponding to outcomes $(a_i,b_i)$ at measurement terminals $i\in[1,4]$ is
	\begin{widetext}
	\begin{align}
		&\left(_{M_{2,1}^1M_{2,2}^1}\bra{\Phi_{a_1,b_1}}~\otimes~_{M_{3,1}^2M_{3,2}^2}\bra{\Phi_{a_2,b_2}}~\otimes~_{M_{4,1}^3M_{4,2}^3}\bra{\Phi_{a_3,b_3}}~\otimes~_{M_{5,1}^4M_{5,2}^4}\bra{\Phi_{a_4,b_4}}\right)\nonumber\\
		&\quad \times\left(\ket{\Psi^+}_{X_1M_{2,1}^1}\otimes\ket{\Psi^+}_{M_{2,2}^1M_{3,1}^2}\otimes\ket{\Psi^+}_{M_{3,2}^2M_{4,1}^3}\otimes\ket{\Psi^+}_{M_{4,2}^3M_{5,1}^4}\otimes\ket{\Psi^+}_{M_{5,2}^4Y_4}\right)\nonumber\\
		&=\left(\frac{1}{2}\right)^4\ket{\Phi_{a_1\oplus a_2\oplus a_3\oplus a_4,b_1\oplus b_2\oplus b_3\oplus b_4}}_{X_1Y_4}.\label{eq-joint_state_post_meas}
	\end{align}
	\end{widetext}
	Therefore, with probability $(\eta^2)^5$, the two branches will share a pair of maximally-entangled photons along the path shown in Fig. \ref{fig-grid}. Since the state \eqref{eq-joint_state_post_meas} can be written in the form \eqref{eq-Bell_states}, it follows that if $X_1$ applies $\sigma_z^{b_1\oplus b_2\oplus b_3\oplus b_4}\sigma_x^{a_1\oplus a_2\oplus a_3\oplus a_4}$ to its photon, then $X_1$ and $Y_4$ will share a pair of photons in the state $\Psi^+$. 
	
	Now, observe using Fig. \ref{fig-grid} that in our protocol all paths from an $X$ branch to a $Y$ branch have the same length. This means that the probability that any two of the $X$ and $Y$ branches share a pair of photons in the state $\Psi^+$ is $(\eta^2)^5$. Note that, so far, we have assumed that the success probability of the Bell measurement is one. If the success probability of the Bell measurement is $\gamma\in[0,1]$, then the probability that any two of the $X$ and $Y$ branches share a pair of photons in the state $\Psi^+$ is $\gamma^4(\eta^2)^5$.
	
	In the general case of $M$ columns of measurement terminals between company $X$ and company $Y$, we find that after a single run of the protocol the state between an $X$ branch and a $Y$ branch along one path of the network is $\ket{\Phi_{a_{\text{tot}},b_{\text{tot}}}}$, where $a_{\text{tot}}=a_1\oplus a_2\oplus\dotsb\oplus a_M$, $b_{\text{tot}}=b_1\oplus b_2\oplus\dotsb\oplus b_M$, and $(a_1,b_1),(a_2,b_2),\dotsc, (a_M,b_M)$ are the measurement outcomes at each measurement terminal. This occurs with probability $\gamma^M(\eta^2)^{M+1}$.

\subsection{Average entanglement yield}\label{subsec-yield}

	Given a network with $N$ branches of company $X$ and company $Y$ and $M$ measurement terminals, at each time step the branches $X_1$, $X_N$, $Y_1$, and $Y_N$ each receive one photon while the rest each receive two photons. The maximum possible number of entangled photon pairs that can be shared between company $X$ and company $Y$ after one time step is therefore $2(N-2)+2=2(N-1)$ and it does not depend on the number of possible paths through which $X$ and $Y$ can be entangled. 
	
	Since the probability of obtaining a single entangled pair of photons between an $X$ branch and a $Y$ branch is $\gamma^M(\eta^2)^{M+1}$, and $\eta=\e^{-\alpha\ell}$, the average total number $\xi_{N,M}^{\text{2D}}$ of entangled pairs created, i.e., the average ``yield'', after one time step is
	\begin{equation}\label{eq-yield_1}
		\xi_{N,M}^{\text{2D}}=2(N-1)\gamma^M \e^{-2\alpha\ell(M+1)}.
	\end{equation}
	In terms of the total horizontal length $L=2\ell(M+1)\cos\theta$ of the network,
	\begin{equation}\label{eq-yield_2}
		\xi_{N,M}^{\text{2D}}=2(N-1)\gamma^M \e^{-\alpha\frac{L}{\cos\theta}}.
	\end{equation}
	
	\begin{figure}
		\centering
			\subfloat[]{\label{subfig-plot_a}\centering\includegraphics[scale=0.4,width=0.22\textwidth]{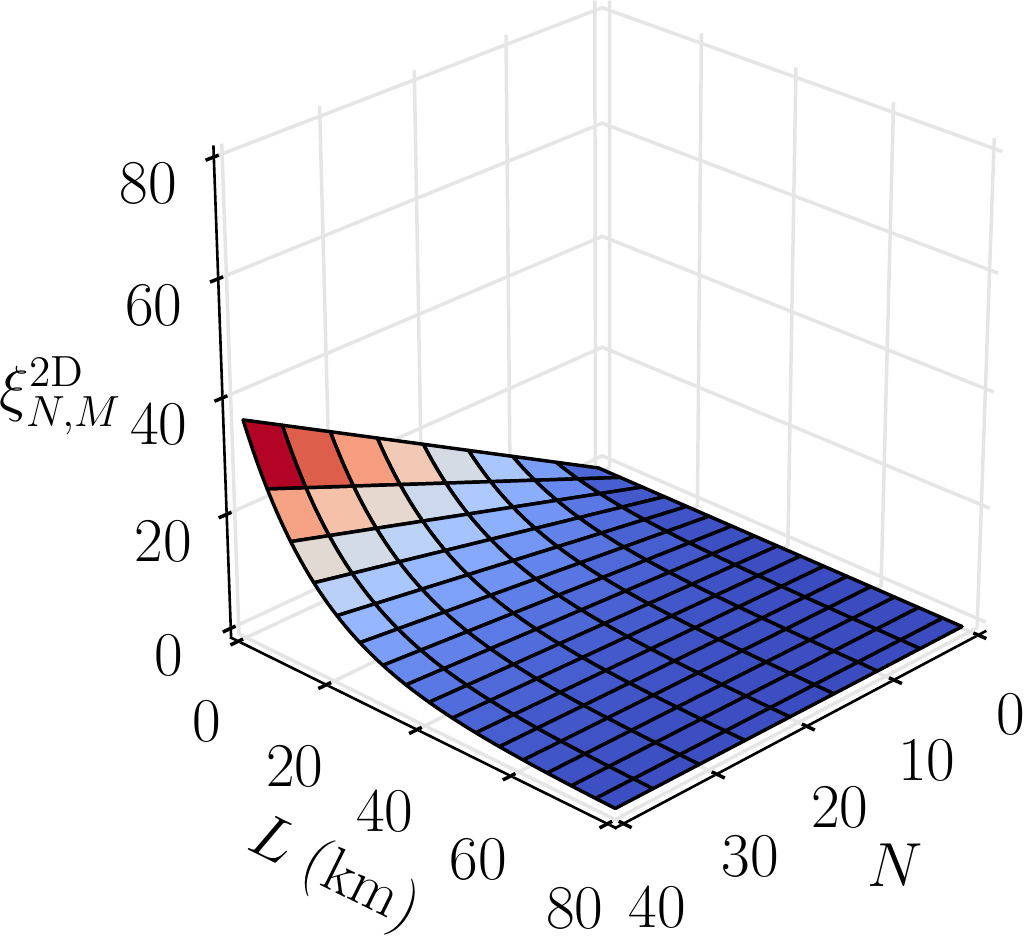}}~
			\subfloat[]{\label{subfig-plot_b}\centering\includegraphics[scale=0.4,width=0.22\textwidth]{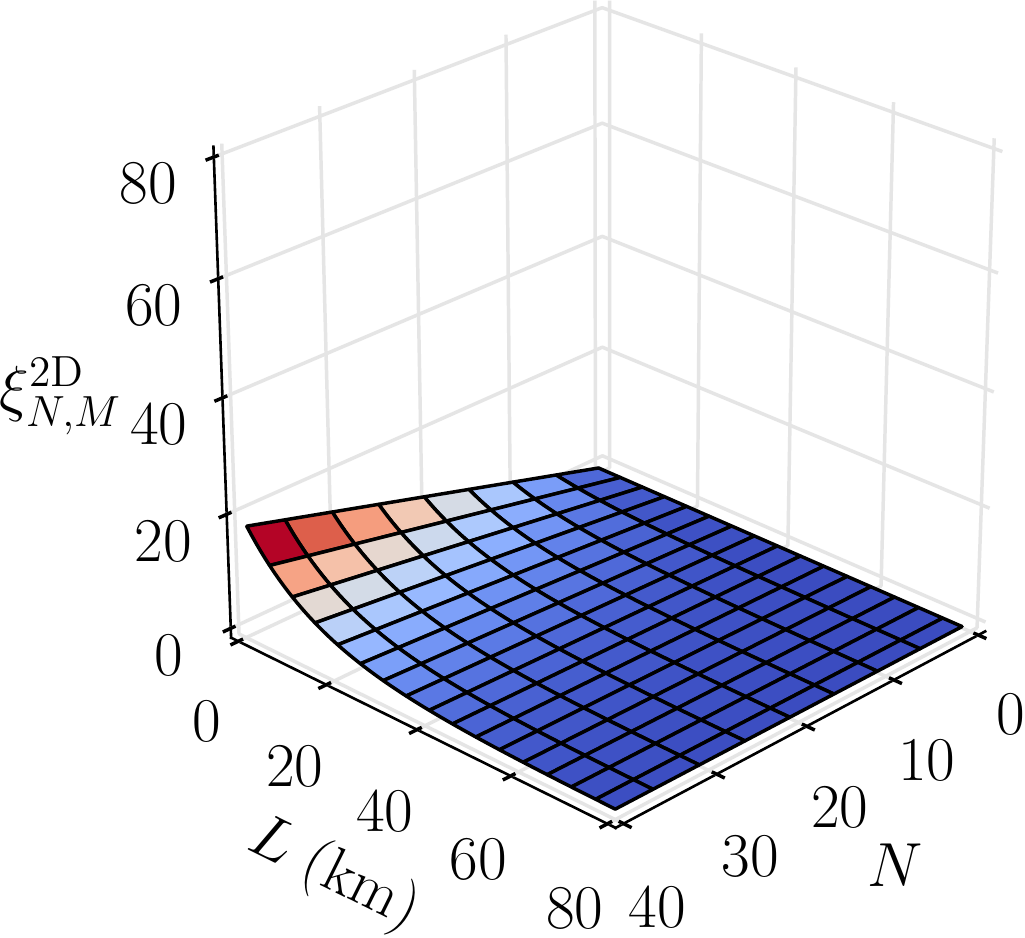}}\\
			\subfloat[]{\label{subfig-plot_c}\centering\includegraphics[scale=0.4,width=0.22\textwidth]{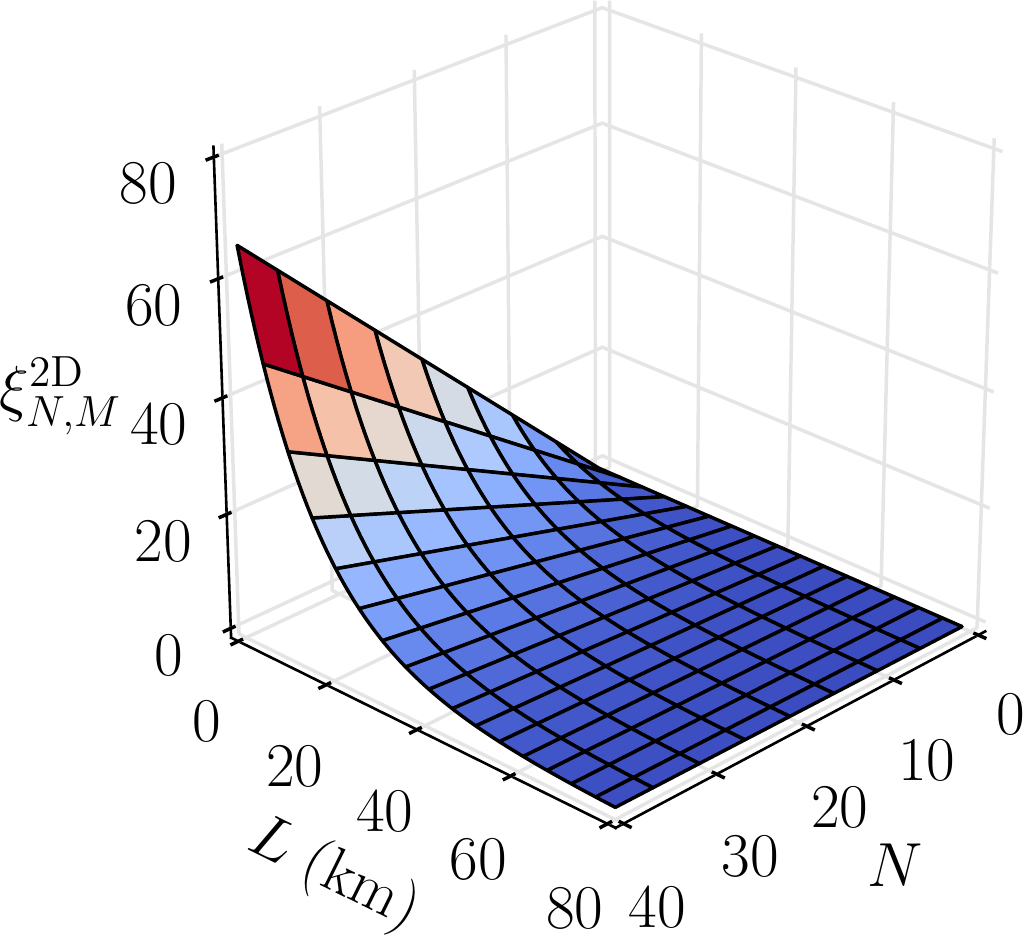}}~
			\subfloat[]{\label{subfig-plot_d}\centering\includegraphics[scale=0.4,width=0.22\textwidth]{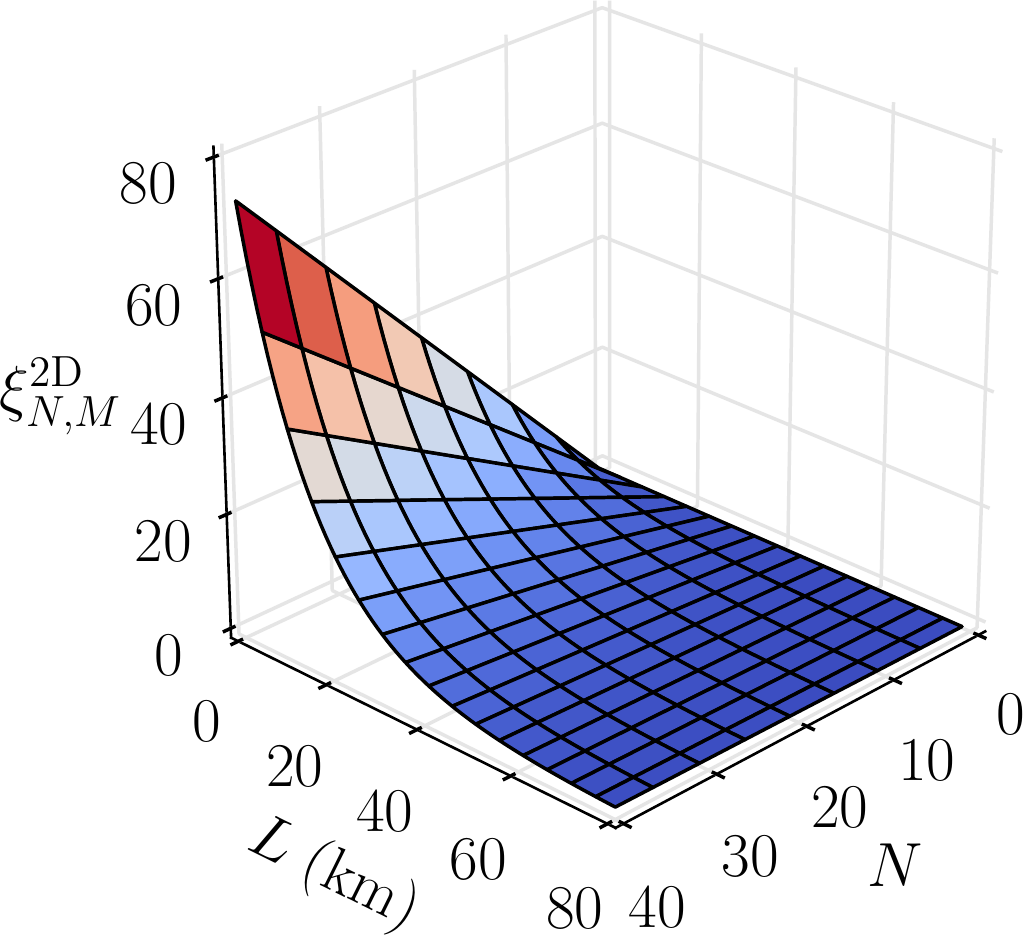}}
		\caption{Plots of $\xi_{N,M}^{\text{2D}}$ as a function of the total length $L$ and the number of branches $N$ for different values of the success probability $\gamma$ of each Bell measurement and $M$, the number of columns of measurement terminals in the network. \subref{subfig-plot_a} $\gamma=\frac{1}{2}$, $M=1$. \subref{subfig-plot_b} $\gamma=\frac{1}{2}$, $M=2$. \subref{subfig-plot_c} $\gamma=0.9$, $M=1$. \subref{subfig-plot_d} $\gamma=1$, $M=2$.}
		\label{fig-yield}
	\end{figure}
	
	We typically let $\alpha=\frac{1}{22\text{ km}}$ \cite{KKL15}. Also, the maximum value of $\gamma$ using linear optics is $\frac{1}{2}$ \cite{VY99,LCS99,CL01}, whereas using non-linear optics a perfect Bell measurement, i.e., $\gamma=1$, is possible in the ideal case \cite{KKS01,KKS02}. Using these values, and letting $\theta=\frac{\pi}{4}$, we plot $\xi_{N,M}^{\text{2D}}$ as a function of the total length $L$ and the number of branches $N$ in Fig. \ref{fig-yield} for $M=1$ and $M=2$. 
	
	\begin{figure}
		\centering
		\subfloat[]{\label{subfig-robust_2}\centering\includegraphics[scale=0.7]{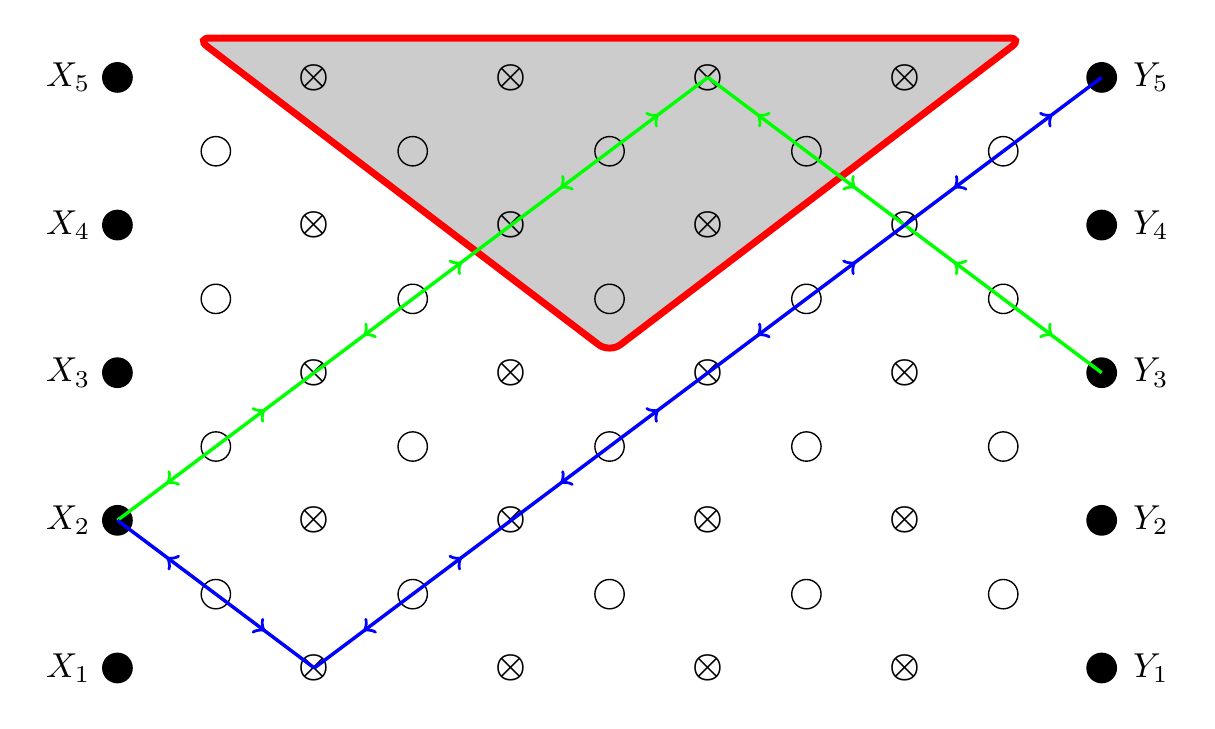}}\\
		\subfloat[]{\label{subfig-robust}\centering\includegraphics[scale=0.7]{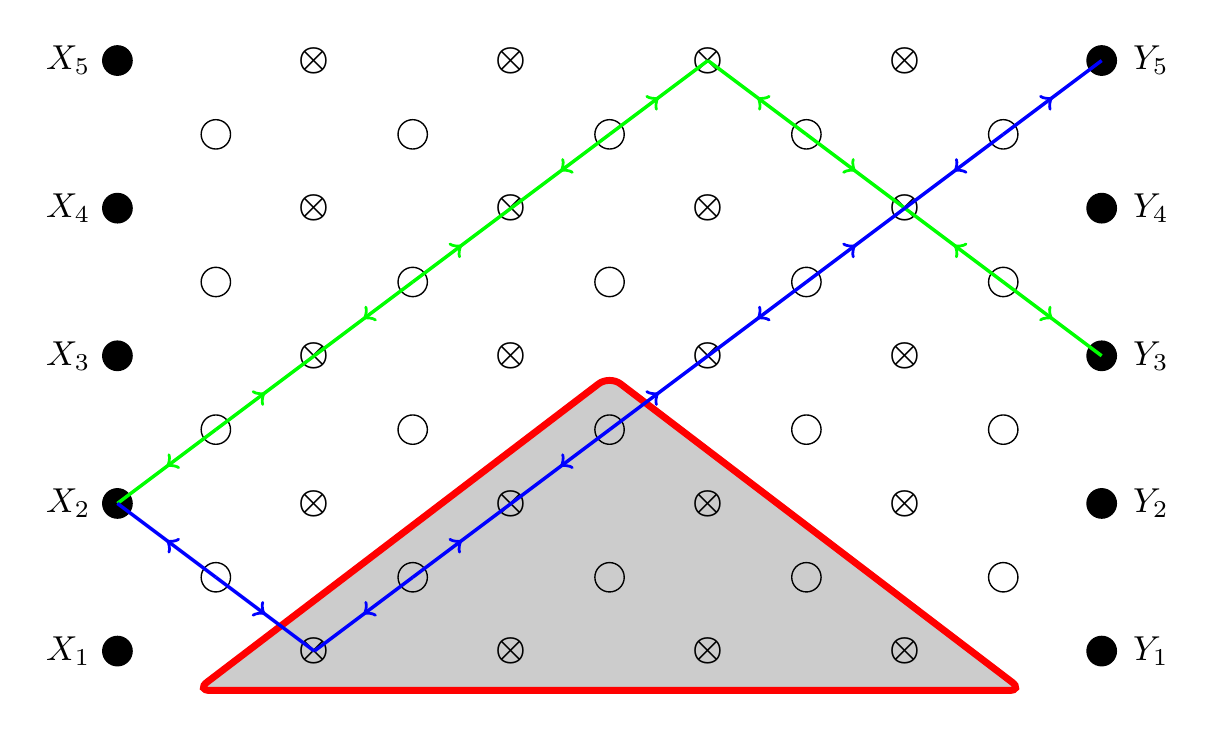}}
		\caption{An illustration of the robustness of the memory-free entanglement distribution protocol presented in Section \ref{subsec:bravais_lattice} in two particular cases of non-functioning source stations and measurement terminals, which are enclosed in the shaded regions. Indicated are two possible paths from $X$ to $Y$. Despite the non-functioning nodes in \subref{subfig-robust_2}, which do not allow shared entanglement between $X_2$ and $Y_3$ via the green path, the blue path is still available to share entanglement between $X_2$ and $Y_5$. Similarly, in \subref{subfig-robust}, the blue path cannot be used to share entanglement, but the green path can be used.}\label{fig-protocol_robust}
	\end{figure}
	
	From \eqref{eq-yield_1} and \eqref{eq-yield_2}, we see that, when viewed as a function of the total length $L$, increasing the number of columns $M$ of measurement terminals in the network has the effect of decreasing the average yield per time step whenever $\gamma<1$, while for $\gamma=1$ the yield is independent of $M$. However, increasing $M$ decreases the distance $\ell$ between the source station and the measurement terminal (for fixed $L$ and $\theta$) and increases the number of paths from one end of the network to the other. Increasing $M$ also increases the number of branches of company $Y$ that a given branch of company $X$ can be entangled with, and vice versa. For example, for the case $N=5$ and $M=4$ in Fig. \ref{fig-grid}, there is no path from $X_1$ to $Y_5$; however, for $M=5$, there exists a path from $X_1$ to $Y_5$. The increase in the number of paths from one end to the other makes the network and protocol robust against failures of intermediate source stations and/or measurement terminals.
	
	An example of the robustness of the protocol against failures of source stations and measurement terminals is illustrated in Fig. \ref{fig-protocol_robust}. The non-functioning nodes are enclosed in the shaded region. The protocol is robust since in each of the two cases of non-functioning nodes as shown in Fig. \ref{subfig-robust_2} and Fig. \ref{subfig-robust}, entanglement can still be distributed between $X$ and $Y$ using the protocol, albeit with a lower average entanglement yield. For example, in Fig. \ref{subfig-robust_2}, the green path cannot be used to allow $X_2$ to share entanglement with $Y_3$, but $X_2$ can still share entanglement with $Y_5$ via the blue path. On the other hand, in Fig. \ref{subfig-robust}, $X_2$ cannot share entanglement with $Y_5$ via the blue path, but it can share entanglement with $Y_3$ via the green path.



\section{Loss tolerance for futuristic network architectures}\label{sec-gen_arc}

	In the previous section, we considered memory-free quantum repeaters in which, at each time step, the Bell measurements at the measurement terminals were performed immediately upon arrival of the photons so as to not require the use of quantum memories to store the qubits until both photons arrive. Ensuring simultaneous arrival of the photons at the measurement terminals and immediate Bell measurement is difficult to achieve in practice \cite{SMI+16}. With technological advancements in the future, we can assume that quantum memories will be easily accessible so that they may be used throughout the network (and not just at select locations) for the storage of the qubits.
	


	In this section, we consider full-fledged 2D network architectures and topologies in which the network is modeled as a graph all of whose nodes represent workstations, i.e., a members of the network that have quantum memories and can perform measurements for entanglement swapping and other quantum operations. In Section \ref{sec-homog}, we model the network as a graph such that the edges connecting two neighboring nodes have the same length, and in Section \ref{sec-inhomog} we consider graphs whose edges have different lengths. In each case, we use percolation theory to quantify the robustness of the networks against photon loss and failures of the nodes.
	
	\begin{figure*}
		\centering
		\includegraphics[scale=0.5]{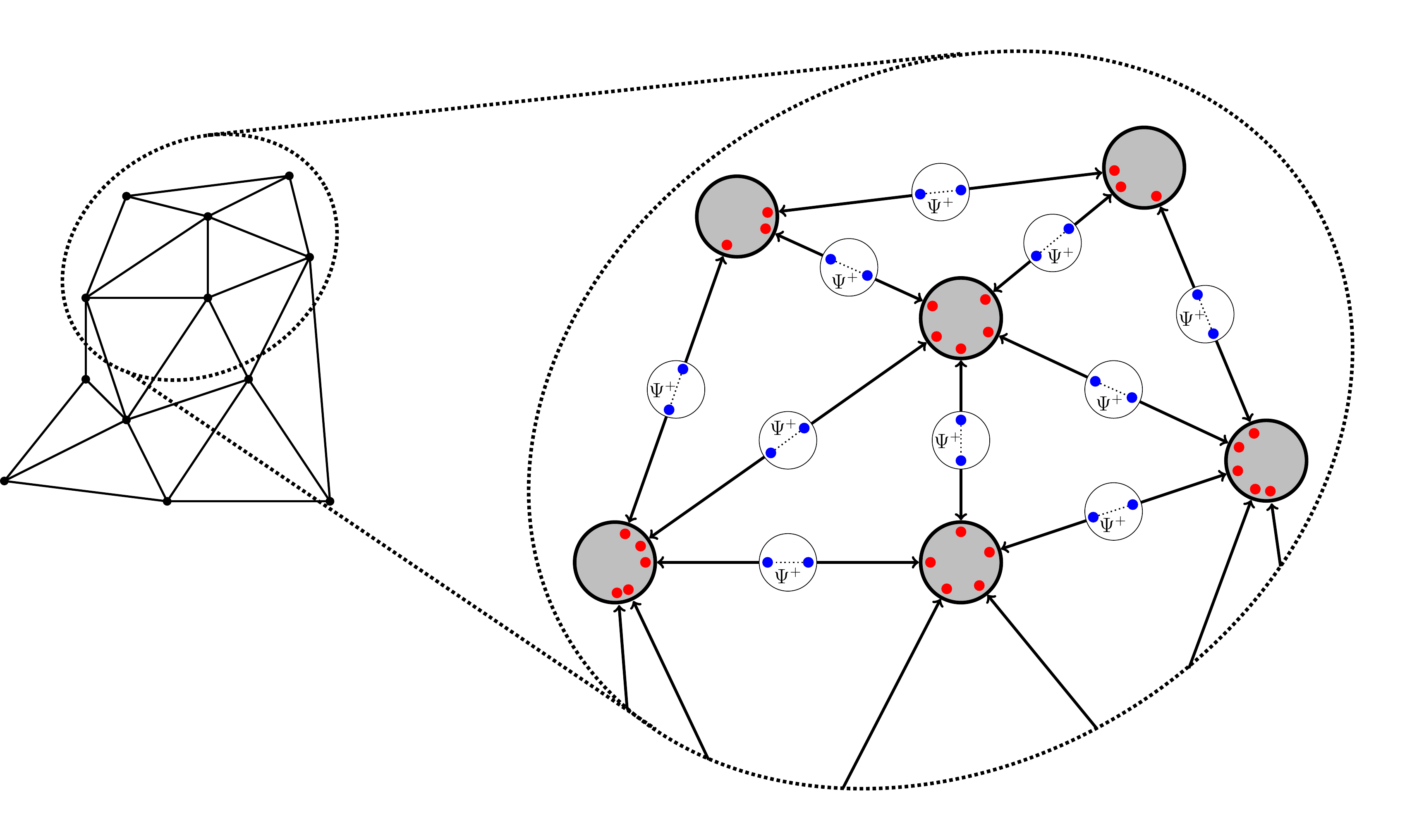}
		\caption{Given a graph (left), the corresponding quantum network architecture is illustrated explicitly for the circled portion of the graph (right) and consists of the following elements: members of the network placed at the nodes of the graph (gray circles), each possessing $d$ quantum memories (represented by red dots inside the nodes), where $d$ is the degree of the node, and source stations (white circles) placed at the midpoint of each edge of the graph that generate pairs of dual-rail single-photonic qubits (represented by blue dots) in the state $\Psi^+$. Each qubit of the pair is fired from the source station in opposite directions along the edge of the graph. The quantum memories allow each member of the network to store the arriving qubits for later processing. Each member of the network has at least one measurement terminal, with a maximum of $\ceil{\frac{d}{2}}$, that can be used to perform Bell measurements for entanglement swapping on any two of the stored qubits.}\label{fig-gen_architecture}
	\end{figure*}
	
	\begin{figure*}
		\centering
			\includegraphics[scale=0.5]{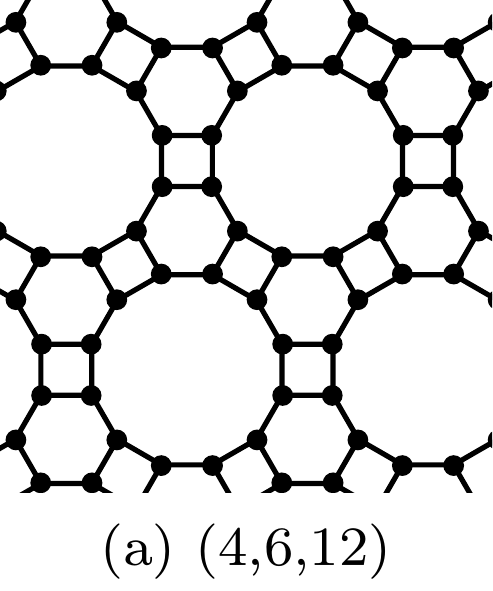}~~
			\includegraphics[scale=0.5]{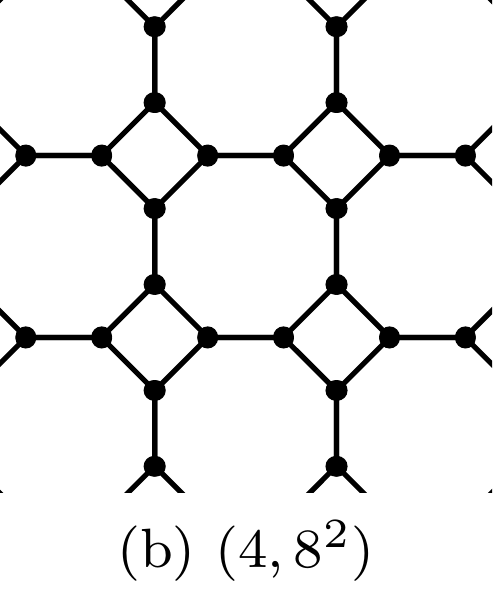}~~
			\includegraphics[scale=0.5]{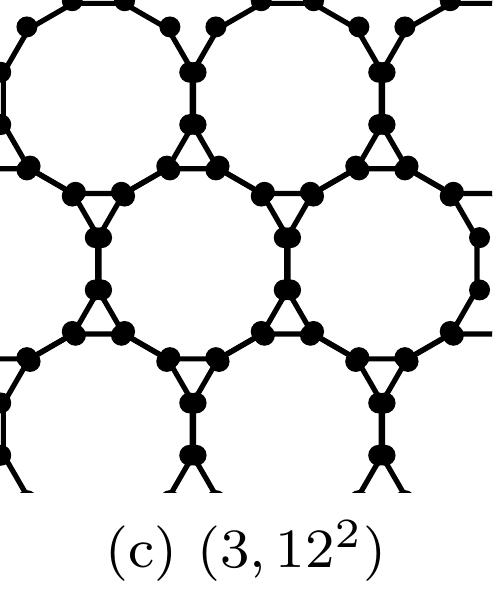}~~
			\includegraphics[scale=0.5]{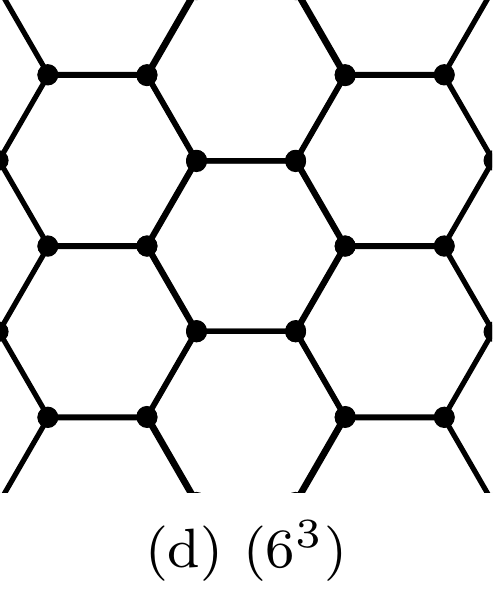}~~
			\includegraphics[scale=0.5]{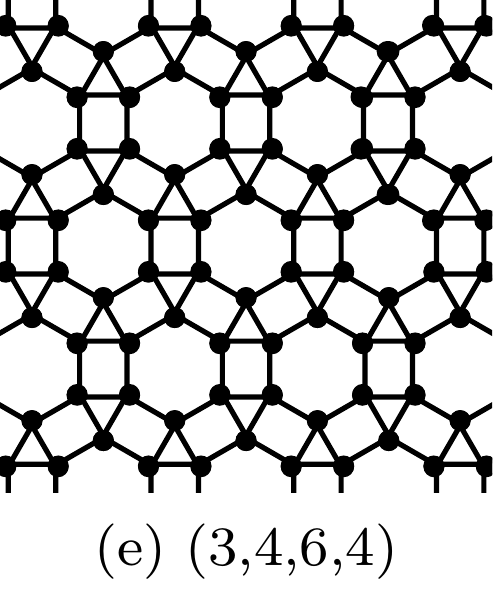}~~
			\includegraphics[scale=0.5]{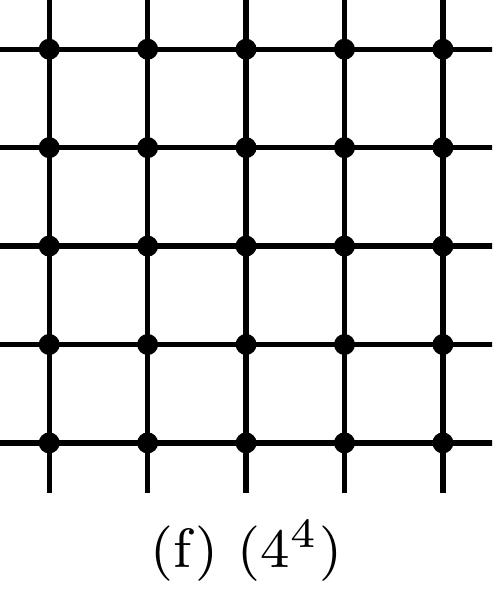}\\[0.5cm]
			\includegraphics[scale=0.5]{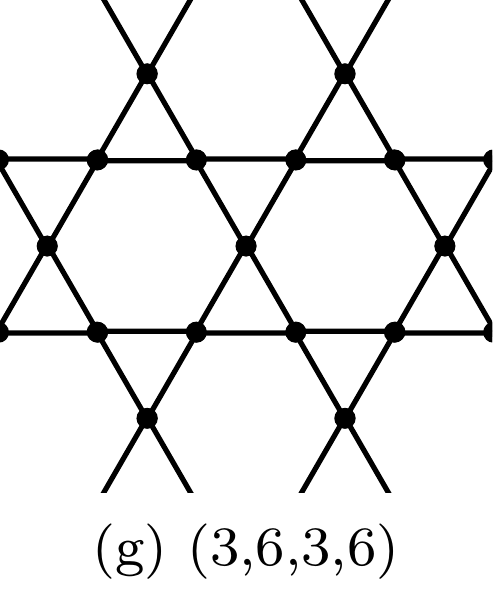}~~
			\includegraphics[scale=0.5]{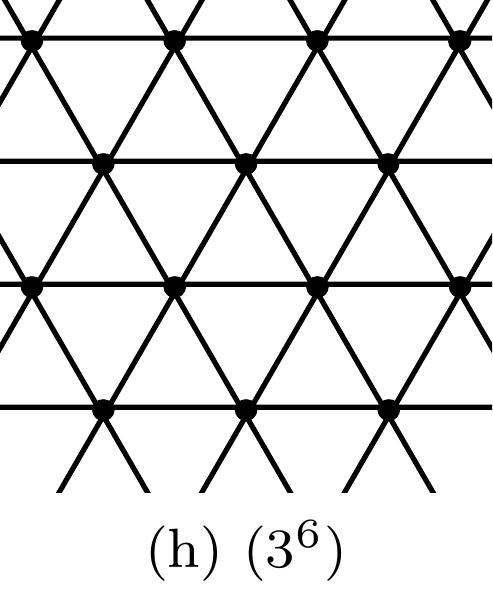}~~
			\includegraphics[scale=0.5]{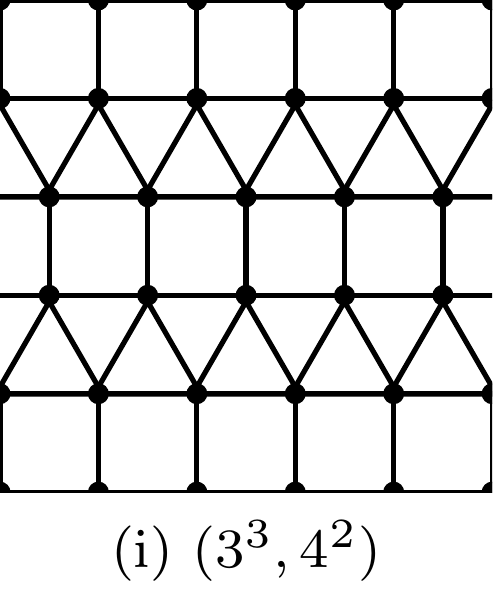}~~
			\includegraphics[scale=0.5]{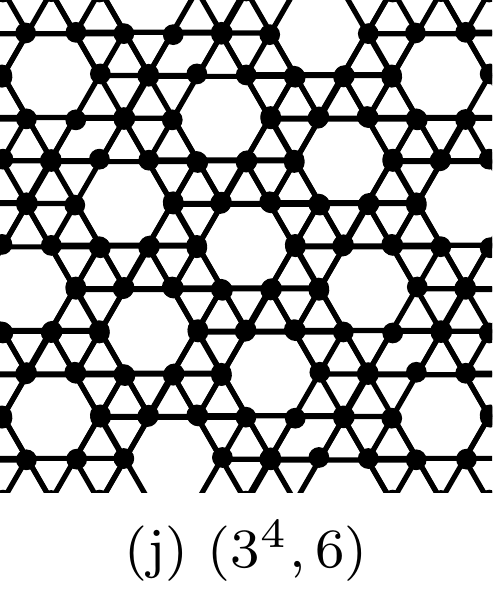}~~
			\includegraphics[scale=0.5]{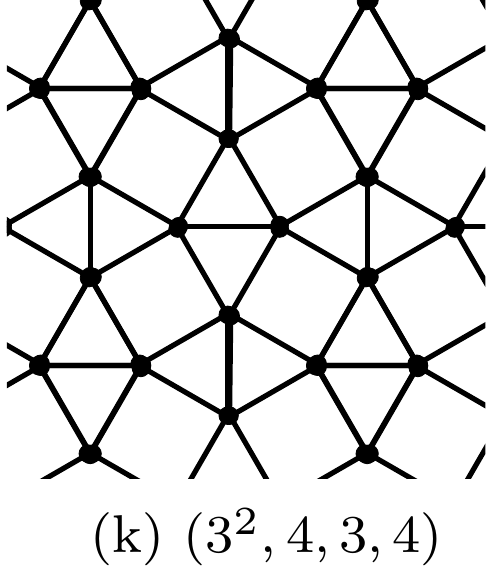}
		\caption{Eleven Archimedean lattices in which all edges of the lattice have the same length. The lattices are named by listing the number of sides of the shapes surrounding each vertex, with repeated shapes indicated with an exponent. For example, $(3^4,6)$ means that every vertex is surrounded by four triangles and one hexagon. Members of the network, capable of storing photons with quantum memories, are represented by $\bullet$. The source stations (not indicated) are placed at the midpoint of each edge between neighboring nodes; see Fig. \ref{fig-gen_architecture}.}
		\label{fig:arch11}
	\end{figure*}
	
	Our generalized network architecture is illustrated in Fig. \ref{fig-gen_architecture}. Given a graph, the corresponding quantum network is defined by placing source stations at the midpoint of each edge of the graph. Each source station generates one pair of dual-rail single-photonic qubits in the Bell state $\Psi^+$ and fires each photon of the pair in opposite directions along the edge towards the neighboring nodes \footnote{It should be noted that one could allow for a different network setup where each workstation is capable of generating entangled pair of photons when needed and entangled photons be stored in its quantum memory or transmitted to the neighboring workstations depending on the task.}. As in Section \ref{sec-ent_dist}, the transmission of the photons along the edge is modeled as a pure-loss bosonic channel. Each node on the graph represents a member of the network that is capable of storing photons using quantum memories and performing Bell measurements for entanglement swapping. The number of quantum memories held by each member of the network is equal to the degree $d$ of the corresponding node in the graph, where the degree of a node in a graph is equal to the number of nodes it is directly connected to through edges. Each member of the network has at least one measurement terminal, with a maximum of $\ceil{\frac{d}{2}}$, for the purpose of entanglement swapping. These measurement terminals can be used on any two of the photons in the quantum memories. Multiple measurement terminals at each node, for example as in Fig. \ref{fig-meas} in which each node contains two measurement terminals, can potentially allow for simultaneous entanglement distribution in the network depending on the protocol used.  
	
	
	
	Depending on the network topology, there may be several possible ways in which any two members of the network can become connected, i.e., share entangled photon pairs. Members of the network, acting in a cooperative manner, can perform entanglement swapping operations (as discussed in Section \ref{sec:ent_swap}) in order to direct the entanglement so that members of interest can become connected. In general, the probability that any two members of the network are connected decreases with the increase in the number of intermediate nodes that participate in the entanglement swapping operations. 
	
\subsection{Homogeneous network topology}\label{sec-homog}
	
\begin{figure}
		\centering
			\subfloat[]{\label{subfig-bowtie_I}\centering\includegraphics[scale=0.5,width=0.20\textwidth]{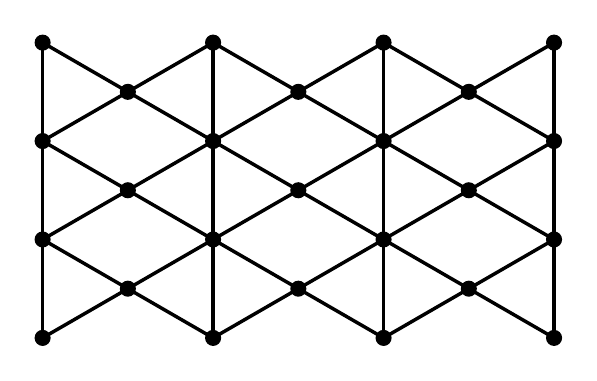}}~
			\subfloat[]{\label{subfig-bowtie_II}\centering\includegraphics[scale=0.5,width=0.20\textwidth]{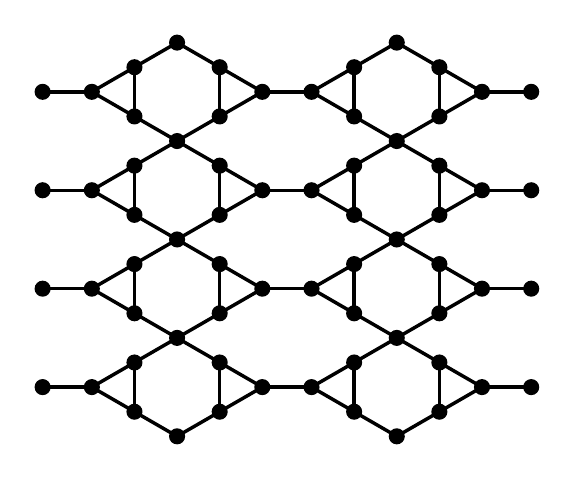}}\\
			\subfloat[]{\label{subfig-bowtie_III}\centering\includegraphics[scale=0.5,width=0.20\textwidth]{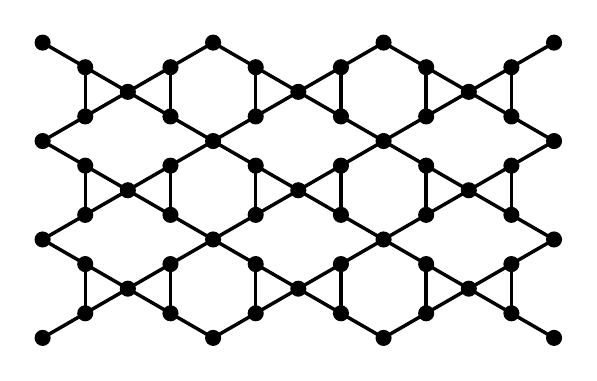}}~
			\subfloat[]{\label{subfig-bowtie_IV}\centering\includegraphics[scale=0.30,width=0.20\textwidth]{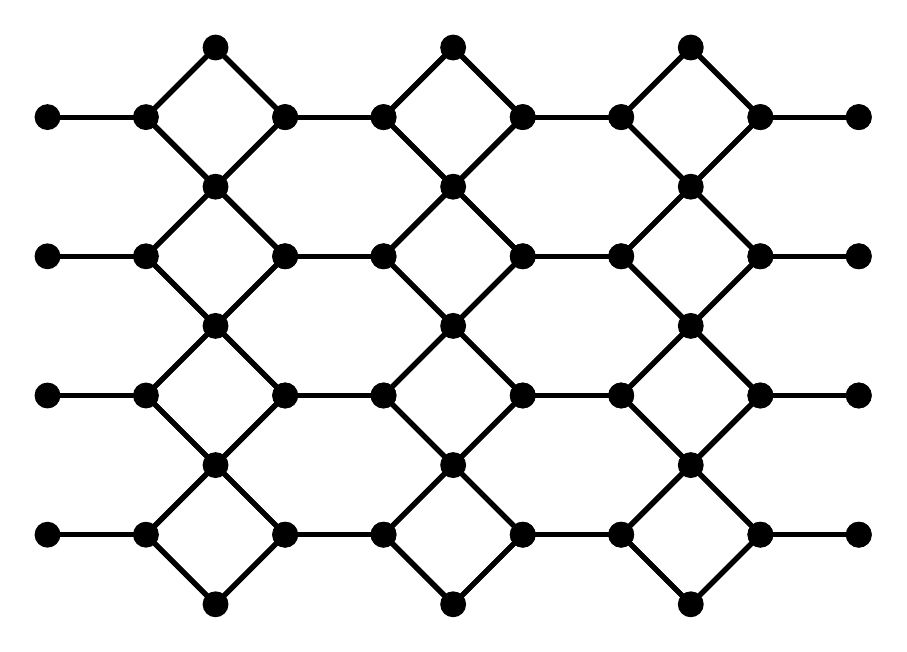}}
		\caption{Four different bow-tie lattices in which all edges of the lattice have the same length. \subref{subfig-bowtie_I} Bow-tie I. \subref{subfig-bowtie_II} Bow-tie II. \subref{subfig-bowtie_III} Bow-tie III. \subref{subfig-bowtie_IV} Bow-tie IV. Members of the network, capable of storing photons with quantum memories, are represented by $\bullet$. The source stations (not indicated) are placed at the midpoint of each edge between neighboring nodes; see Fig. \ref{fig-gen_architecture}.}
		\label{fig-bowtie}
	\end{figure}

	A homogeneous network is one in which all edges of the corresponding graph have the same length. Graphs whose edges all have the same length include the four bow-tie lattices in Fig. \ref{fig-bowtie} and the 11 Archimedean lattices in Fig. \ref{fig:arch11}. As described above, in these two figures each member of the network is represented by $\bullet$, and the source stations (not indicated) are placed at the midpoint of each edge between neighboring nodes. Members of the network are capable of storing photons with quantum memories and performing entanglement swapping operations by Bell measurements. By the calculation in Section \ref{sec-ent_dist}, any two neighboring nodes share the entangled state $\Psi^+$ with probability $\eta^2$, and with probability $1-\eta^2$ at least one of the photons is lost and there is no entanglement between the nodes. 
	
	Of interest in a quantum network is the ability to establish long-range connections between any two nodes in the network. Whether such long-range connections are possible in the case when entanglement between neighboring nodes is established probabilistically along the edges as described above can be answered using percolation theory (see, e.g., Ref. \cite{Gri99book}), specifically bond percolation theory \footnote{Previous studies of entanglement distribution in large networks based on percolation theory can be found in Refs. \cite{ACL07,PCA+08,CC09,LWL09}.}.
	
	In bond percolation theory, any two neighboring nodes of a given graph are either connected with probability $p$ or disconnected with probability $1-p$. One of the central questions of percolation theory is whether there exists a giant cluster of connected nodes in the graph such that a path connects one end of the graph to the other. It turns out that there is a critical probability $p_c$ above which such a cluster always exists. In general, the critical probability can be determined numerically (see, e.g., Refs. \cite{SZ99,NZ01}), while for certain classes of graphs the critical probability can be determined analytically (see, e.g., Refs. \cite{W84,ZS06}).
	
\begin{figure}
		\centering
		\includegraphics[scale=1]{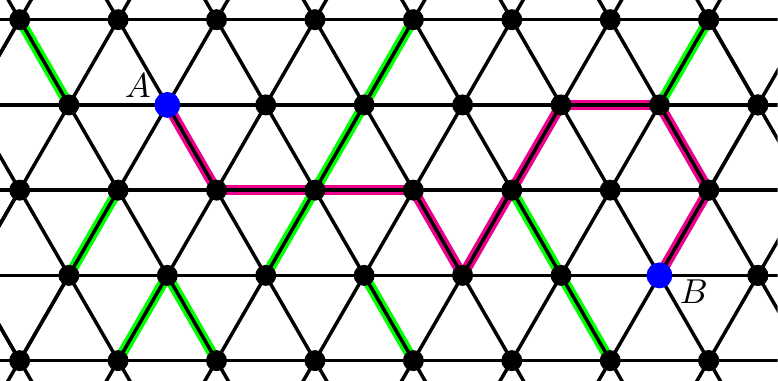}
		\caption{In this quantum network based on the $(3^6)$ Archimedean lattice there exists a path, i.e., a chain of successfully-entangled nodes (indicated in magenta), between the two distant nodes $A$ and $B$ (indicated in blue). Through entanglement swapping at the intermediate nodes along the path, $A$ and $B$ can share entanglement. Due to the probabilistic nature of entanglement generation in our architecture, at any given time, sources firing throughout the network can lead to paths between multiple different pairs of nodes. Some of these other paths are indicated in green.}\label{fig-perc}
	\end{figure}

	Now, for the network architectures considered here, we deem two neighboring nodes to be connected if they share the entangled state $\Psi^+$, i.e., if the transmission of the entangled pair of photons in the state $\Psi^+$ from the source station at the midpoint of the edge connecting the two nodes succeeds. The probability that any two neighboring nodes are connected is therefore $\eta^2$. Since all members of the network have quantum memories, unlike the protocol in Section \ref{subsec:bravais_lattice}, once the connection has been established  they are not forced to measure immediately upon receiving the photons and therefore can hold on to their half of the entangled pair of photons for later processing. As illustrated in Fig. \ref{fig-perc} for the $(3^6)$ Archimedean lattice, a path between two (potentially distant) nodes of interest (indicated in blue) constitutes a chain of connected pairs of nodes (indicated in magenta) between the two given nodes. If there exists a path between two nodes of interest, then entanglement swapping operations performed at the intermediate nodes along the path can be used to establish entanglement between the two nodes. Now, it is possible that, at any given time, sources firing throughout the network can lead to multiple paths between the two nodes of interest, and even paths between multiple different pairs of nodes. These other paths are indicated in green in Fig. \ref{fig-perc}, and in general they can allow for simultaneous sharing of entanglement between multiple different pairs of nodes in the network.
	

	We define the critical transmissivity $\eta_c$ for a given graph as the transmissivity of the pure-loss bosonic channel above which there exists a giant cluster of connected nodes in the corresponding network such that entanglement can be established between one end of the network and the other by entanglement swapping at intermediate nodes. Given that, in our noise model, entanglement is established along an edge between neighboring nodes with probability $\eta^2$, the critical transmissivity $\eta_c$ for a given graph is simply the square root of the critical probability $p_c^{\text{bond}}$ for bond percolation \footnote{More generally, for any noise model in a homogeneous network topology, if $\beta$ is the probability of sharing perfect entanglement, i.e., a Bell pair, between any two neighboring nodes, then the critical probability $\beta_c$ of establishing a large cluster of entangled nodes is equal to $p_c^{\text{bond}}$. Note that subsequent entanglement swapping operations to establish entanglement between two distant nodes in such a cluster may have to be supplemented with purification protocols.}, i.e.,
	\begin{equation}\label{eq-trans_crit}
		\eta_c=\sqrt{p_c^{\text{bond}}}.
	\end{equation} 
	We deem the network robust against photon loss whenever $\eta\geq\eta_c$.
	
	Now, suppose that in addition to the probability $\eta^2$ of establishing entanglement between neighboring nodes along the edges of the graph the fiber optic cable connecting the neighboring nodes malfunctions with probability $q$. Then, the overall probability of establishing entanglement between neighboring nodes along an edge is $\eta^2(1-q)$. By comparing this probability with $p_c^{\text{bond}}$, one can find values of $\eta$ and $q$ such that a giant cluster of entangled nodes exists in the network despite failures of the fiber optic cables. Specifically, the condition 
	\begin{equation}\label{eq-robust_region}
		\eta^2(1-q)\geq p_c^{\text{bond}}
	\end{equation}
	defines the \textit{region of robustness} of the network as the values of $\eta$ and $q$ for the which the inequality \eqref{eq-robust_region} is satisfied.
	
	
	It may also happen that at any given time some fraction of the workstations malfunctions. If we suppose that each workstation is well-functioning with probability $r$ and malfunctions with probability $1-r$, then the question of whether a giant cluster of entangled nodes exists in the network can be answered using site percolation theory. As opposed to bond percolation theory, in site percolation nodes are either present with probability $p$ or absent with probability $1-p$. The critical site percolation probability $p_c^{\text{site}}$ is defined as the value above which a giant cluster of connected nodes exists in the network. The critical value $r_c$ of $r$ above which there exists a giant cluster of entangled nodes is therefore simply $r_c=p^{\text{site}}_c$, assuming $\eta=1$ and $q=0$. Table \ref{table-crit_bowtie} and Table \ref{table-crit_arch} list the critical probabilities for bond and site percolation, as well as the critical transmissivities, for the bow-tie and Archimedean lattices, respectively.
	
	
	\begin{table}
		\centering
		\begin{tabular}{|c||c|c|c|}
			\hline Bow-tie lattice & $p_c^{\text{bond}}$ & $r_c=p_c^{\text{site}}$ & $\eta_c$ \\ \hline\hline
			I & 0.404518 & 0.5475 \cite{VDM97} & 0.636017 \\ \hline
			II & 0.672929 & unknown & 0.820322 \\ \hline
			III & 0.625457 & unknown & 0.790858 \\ \hline
			IV & 0.595482 & unknown & 0.771674 \\ \hline 
		\end{tabular}
		\caption{Bond percolation critical probabilities for the bow-tie lattices in Fig. \ref{fig-bowtie} as determined in Ref. \cite{ZS06} along with the critical transmissivity as given by \eqref{eq-trans_crit}. Site percolation critical probabilities for these lattices are unknown except for the bow-tie I lattice.}
		\label{table-crit_bowtie}
	\end{table}

	\begin{table}
		\centering
		\begin{tabular}{|>{\centering\arraybackslash}m{2cm}||>{\centering\arraybackslash}m{2cm}|>{\centering\arraybackslash}m{2cm}|>{\centering\arraybackslash}m{1.5cm}|}
			\hline \begin{tabular}{c} Archimedean\\lattice\end{tabular} & $p_c^{\text{bond}}$ & $r_c=p_c^{\text{site}}$ & $\eta_c$ \\ \hline\hline
			$(3,12^2)$ & 0.740421 \cite{P07} & 0.807904 \cite{SZ99} & 0.860477 \\ \hline
			$(4,6,12)$ & 0.693733 \cite{P07} & 0.747806 \cite{SZ99} & 0.832906 \\ \hline
			$(4,8^2)$  & 0.676802 \cite{P07} & 0.729724 \cite{SZ99} & 0.822679 \\ \hline
			$(6^3)$ & 0.652703 \cite{SE64} & 0.697043 \cite{SZ99} & 0.807900 \\ \hline
			$(3,6,3,6)$ & 0.524404 \cite{Jac14} & 0.652703 \cite{SE64} & 0.724157 \\ \hline
			$(3,4,6,4)$ & 0.524832 \cite{P07} & 0.621819 \cite{SZ99} & 0.724452 \\ \hline
			$(4^4)$ & $\frac{1}{2}$ & 0.592746 \cite{Jac14} & 0.707106 \\ \hline
			$(3^4,6)$ & 0.434306 \cite{P07} & 0.579498 \cite{SZ99} & 0.659018 \\ \hline
			$(3^3,4^2)$ & 0.419641 \cite{P07} & 0.550213 \cite{SZ99} & 0.647797 \\ \hline
			$(3^2,4,3,4)$ & 0.414137 \cite{P07} & 0.550806 \cite{SZ99} & 0.643534 \\ \hline
			$(3^6)$ & 0.347296 \cite{SE64} & $\frac{1}{2}$ & 0.589318  \\ \hline
		\end{tabular}
		\caption{Bond and site percolation critical probabilities for the Archimedean lattices in Fig. \ref{fig:arch11} along with the critical transmissivity as given by \eqref{eq-trans_crit}.}
		\label{table-crit_arch}
	\end{table}
	
	The critical transmissivity $\eta_c$ and the critical site probability $r_c$ are figures of merit for characterizing the robustness of a network. Networks with lower values of $\eta_c$ and $r_c$ are more fail-safe than those with higher values, because networks with lower values of $\eta_c$ and $r_c$ contain a giant cluster of entangled nodes despite the high probability of photon loss and/or high probability of workstation failure. Among the bow-tie and Archimedean lattices shown above, we find by examining Table \ref{table-crit_bowtie} and Table \ref{table-crit_arch} that the triangular $(3^6)$ Archimedean lattice has the lowest bond percolation critical probability, which is $0.347296$, with $\eta_c=0.589318$. This value of the transmissivity is of practical interest \cite{VanMeter_book}. The triangular lattice also has the lowest site percolation critical probability, which is $\frac{1}{2}$.   
	
\subsection{Inhomogeneous network topology}\label{sec-inhomog}
	
	The results above can be generalized to the case when not all of the lengths of the edges of the graph are the same. This generalization corresponds to different entanglement generation probabilities along the edges, and the question of whether a large cluster of connected nodes exists in the network can be answered using inhomogeneous bond percolation theory. In inhomogeneous bond percolation on regular lattices, each edge of the unit cell comprising the lattice has a different connection probability with a neighboring node. Instead of a critical probability, one obtains in this case a critical surface defining the region of the different probabilities in which a large cluster of connected nodes exists.
	
	Exact critical surfaces exist for the square $(4^4)$, triangular $(3^6)$, and honeycomb $(6^3)$ Archimedean lattices \cite{SE64}, while approximate or conjectured exact critical surfaces exist for many of the other Archimedean lattices \cite{SZ06,SZ08,SZ10}. Conjectured critical surfaces for the bow-tie lattices can be found in Refs. \cite{SZ08,ZSWS12}.
	
	Fig. \ref{fig-inhomog_percolation} shows the unit cells for the square, triangular, honeycomb, and bow-tie I lattices along with their inhomogeneous bond percolation edge connection probabilities. The associated bond percolation critical surfaces are as follows:

	\begin{figure}
		\centering
			\subfloat[]{\label{subfig-square_unit}\centering\includegraphics[scale=1,width=0.11\textwidth]{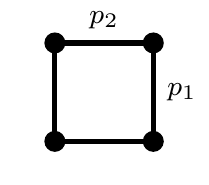}}~~
			\subfloat[]{\label{subfig-triangle_unit}\centering\includegraphics[scale=1,width=0.08\textwidth]{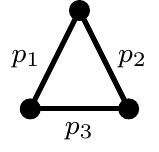}}~~
			\subfloat[]{\label{subfig-honeycomb_unit}\centering\includegraphics[scale=1,width=0.10\textwidth]{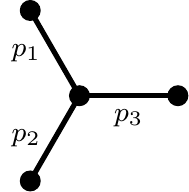}}~~
			\subfloat[]{\label{subfig-bowtie_a_unit}\centering\includegraphics[scale=1,width=0.08\textwidth]{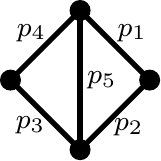}}
		\caption{Unit cell and inhomogeneous bond percolation connection probabilities for: \subref{subfig-square_unit} the square $(4^4)$ Archimedean lattice; \subref{subfig-triangle_unit} the triangular $(3^6)$ Archimedean lattice; \subref{subfig-honeycomb_unit} the honeycomb $(6^3)$ Archimedean lattice; and \subref{subfig-bowtie_a_unit} the bow-tie I lattice.}
		\label{fig-inhomog_percolation}
	\end{figure}
	
	\begin{align}
		\text{Square}~(4^4)~\text{\cite{SE64}} & ~~ p_1+p_2=1,\\
		\text{Triangular}~(3^6)~\text{\cite{SE64}} & ~~ p_1+p_2+p_3-p_1p_2p_3=1, \label{eq-crit_surf_triangle}\\
		\text{Honeycomb}~(6^3)~\text{\cite{SE64}} & ~~ p_1p_2+p_1p_3 \nonumber \\
									   & ~~ +p_2p_3-p_1p_2p_3=1,\\
		\text{Bow-tie I}~\text{\cite{SZ08}} & ~~ p_5+p_1p_2+p_1p_3+p_1p_4 \nonumber \\
							 &~~ +p_2p_3+p_2p_4+p_3p_4 \nonumber \\
							 &~~ -p_1p_2p_3-p_1p_2p_4-p_1p_3p_4 \nonumber \\
							 &~~ -p_2p_3p_4-p_1p_2p_5+p_3p_4p_5 \nonumber \\
							 &~~ -p_1p_2p_3p_4p_5=1.
	\end{align}
	
	Here, $p_i=\eta_i^2\coloneqq \e^{-\alpha\ell_i}$, where $\ell_i$ is the length of the edge along which the probability of establishing entanglement is $\eta_i^2$. These critical surfaces can be used to determine, for a given $\alpha$, critical values of the lengths of the edges for the existence of a large cluster of entangled nodes in the network. 
	
	Though increasing the lengths of the some of the edges in the unit cell can allow for entanglement distribution over longer distances, doing so will generally increase the critical transmissivity, as one might expect. For example, suppose we take a right isosceles triangle as the unit cell of the triangular lattice, so that the length of the two equal sides is $\ell$ and the length of the hypotenuse is $\sqrt{2}\ell$. Then $p_1=p_2\equiv p$ and $p_3=p^{\sqrt{2}}$, and by \eqref{eq-crit_surf_triangle} the critical probability $p_c$ of the graph satisfies $2p_c+p_c^{\sqrt{2}}-p_c^{2+\sqrt{2}}=1$, leading to $p_c\approx 0.388510$, which is larger than the bond percolation probability on the regular triangular lattice of $0.347296$. This suggests that networks based on homogeneous lattices are more robust than those based on inhomogeneous lattices.
	



\section{Discussion}\label{sec-discussion}
	
	
	In Section \ref{subsec:bravais_lattice}, we presented a two-dimensional quantum network architecture consisting of source stations producing entangled photon pairs and measurement terminals performing entanglement swapping Bell measurements without the use of quantum memories. We then used this network architecture to present a protocol for simultaneous entanglement generation among members of two groups $X$ and $Y$ at opposite ends of the network in which the intermediate source stations and measurement terminals all act independently of each other. As discussed in Section \ref{subsec-yield}, this protocol is robust against failures of many of these intermediate nodes. To enhance the robustness of the protocol, the protocol can be modified so that the intermediate nodes do not act independently. For example, the source stations can, upon learning of the failures of the neighboring source stations, modify the configuration of their entanglement generation to one of the three different ways shown in Fig. \ref{fig-sources}. Similarly, the measurement terminals can measure in the different orientations as shown in Fig. \ref{fig-meas} based on the failures of the neighboring measurement terminals. This adaptive approach could allow for paths that would otherwise be disallowed due to the failures to be redirected through active nodes and thus remain viable for sharing entanglement between the $X$ and $Y$ branches. Such a strategy based on using the knowledge of neighboring nodes has also been shown to improve the average entanglement yield, see Ref. \cite{PKT+17}. 
	
	More generally, the network architecture of Section \ref{subsec:bravais_lattice} can be modified so that the intermediate measurement terminals are also members of the network and all members have quantum memories. In this case, individual members can act as routers by selectively measuring in the different orientations as shown in Fig. \ref{fig-meas}. The source stations can also generate entanglement in all three different ways as shown in Fig. \ref{fig-sources}. These generalizations will increase the number of paths between any two nodes in the network. Members of the network can then, by acting cooperatively, exploit this multitude of paths in the network to generate entanglement between any two members of the network. By making full use of the network architecture in this way, the members of the network can execute a variety of different protocols, not just the one presented in Section \ref{subsec:bravais_lattice}. These considerations also apply to the network architectures presented in Section \ref{sec-gen_arc}. The directing of entanglement swapping, in the intermediate workstations through different paths, to generate entanglement between the members of a network, is called entanglement routing. For general discussions on entanglement routing, see, e.g., Refs. \cite{SMI+16,P17,PKT+17}. We emphasize that though the number of paths through any given network may be high, in any single run of a protocol only a certain fraction of these paths might actually be available to distribute entanglement. The number of paths available will depend on the number of photons distributed to the nodes during the protocol. 
	


	The network architectures presented here can also be used for the distribution of entanglement between atoms \cite{DLCZ01,DK03,BK05,DHR17} via the interaction of the photons resulting from transitions in the states of the atom. For example, the network architecture in Section \ref{subsec:bravais_lattice} can be ``inverted'', so that the source stations in Fig. \ref{fig-grid} are replaced by measurement terminals and the measurement terminals are replaced by quantum memories holding matter qubits that act as photon sources when the atoms in the memory undergo state transitions. This inverted network architecture can be used, for example, to perform quantum clock synchronization \cite{JADW00,ITDB17} and other quantum information processing tasks requiring matter entanglement. 
	
	For future work, it would be interesting to adapt the network architectures we have considered in this work to the generation of multipartite entangled states \cite{WZM+16,EKB16,EKB16b}, such as multipartite GHZ states or more generally graph states \cite{BR01}.


\begin{acknowledgments}

	The authors thank Norbert L\"{u}tkenhaus, Rodney van Meter, and Mark M. Wilde for helpful discussions. SD acknowledges support from the LSU Graduate School Economic Development Assistantship. SK acknowledges support from the LSU Department of Physics and Astronomy. JPD would like to acknowledge support from the Army Research Office, Air Force Office of Scientific Research, Defense Advanced Research Projects Agency, National Science Foundation and Northrop Grumman Aerospace Systems.
	
\end{acknowledgments}

SD and SK contributed equally to this work.

\bibliography{kdd}{}

\begin{thebibliography}{74}%
\makeatletter
\providecommand \@ifxundefined [1]{%
 \@ifx{#1\undefined}
}%
\providecommand \@ifnum [1]{%
 \ifnum #1\expandafter \@firstoftwo
 \else \expandafter \@secondoftwo
 \fi
}%
\providecommand \@ifx [1]{%
 \ifx #1\expandafter \@firstoftwo
 \else \expandafter \@secondoftwo
 \fi
}%
\providecommand \natexlab [1]{#1}%
\providecommand \enquote  [1]{``#1''}%
\providecommand \bibnamefont  [1]{#1}%
\providecommand \bibfnamefont [1]{#1}%
\providecommand \citenamefont [1]{#1}%
\providecommand \href@noop [0]{\@secondoftwo}%
\providecommand \href [0]{\begingroup \@sanitize@url \@href}%
\providecommand \@href[1]{\@@startlink{#1}\@@href}%
\providecommand \@@href[1]{\endgroup#1\@@endlink}%
\providecommand \@sanitize@url [0]{\catcode `\\12\catcode `\$12\catcode
  `\&12\catcode `\#12\catcode `\^12\catcode `\_12\catcode `\%12\relax}%
\providecommand \@@startlink[1]{}%
\providecommand \@@endlink[0]{}%
\providecommand \url  [0]{\begingroup\@sanitize@url \@url }%
\providecommand \@url [1]{\endgroup\@href {#1}{\urlprefix }}%
\providecommand \urlprefix  [0]{URL }%
\providecommand \Eprint [0]{\href }%
\providecommand \doibase [0]{http://dx.doi.org/}%
\providecommand \selectlanguage [0]{\@gobble}%
\providecommand \bibinfo  [0]{\@secondoftwo}%
\providecommand \bibfield  [0]{\@secondoftwo}%
\providecommand \translation [1]{[#1]}%
\providecommand \BibitemOpen [0]{}%
\providecommand \bibitemStop [0]{}%
\providecommand \bibitemNoStop [0]{.\EOS\space}%
\providecommand \EOS [0]{\spacefactor3000\relax}%
\providecommand \BibitemShut  [1]{\csname bibitem#1\endcsname}%
\let\auto@bib@innerbib\@empty
\bibitem [{\citenamefont {Kimble}(2008)}]{Kim08}%
  \BibitemOpen
  \bibfield  {author} {\bibinfo {author} {\bibfnamefont {H.~J.}\ \bibnamefont
  {Kimble}},\ }\href {http://dx.doi.org/10.1038/nature07127} {\bibfield
  {journal} {\bibinfo  {journal} {Nature}\ }\textbf {\bibinfo {volume} {453}}
  (\bibinfo {year} {2008})}\BibitemShut {NoStop}%
\bibitem [{\citenamefont {Bennett}\ \emph {et~al.}(1993)\citenamefont
  {Bennett}, \citenamefont {Brassard}, \citenamefont {Cr\'epeau}, \citenamefont
  {Jozsa}, \citenamefont {Peres},\ and\ \citenamefont {Wootters}}]{BBC+93}%
  \BibitemOpen
  \bibfield  {author} {\bibinfo {author} {\bibfnamefont {C.~H.}\ \bibnamefont
  {Bennett}}, \bibinfo {author} {\bibfnamefont {G.}~\bibnamefont {Brassard}},
  \bibinfo {author} {\bibfnamefont {C.}~\bibnamefont {Cr\'epeau}}, \bibinfo
  {author} {\bibfnamefont {R.}~\bibnamefont {Jozsa}}, \bibinfo {author}
  {\bibfnamefont {A.}~\bibnamefont {Peres}}, \ and\ \bibinfo {author}
  {\bibfnamefont {W.~K.}\ \bibnamefont {Wootters}},\ }\href {\doibase
  10.1103/PhysRevLett.70.1895} {\bibfield  {journal} {\bibinfo  {journal}
  {Physical Review Letters}\ }\textbf {\bibinfo {volume} {70}},\ \bibinfo
  {pages} {1895} (\bibinfo {year} {1993})}\BibitemShut {NoStop}%
\bibitem [{\citenamefont {Braunstein}\ \emph {et~al.}(2000)\citenamefont
  {Braunstein}, \citenamefont {Fuchs},\ and\ \citenamefont {Kimble}}]{BFK00}%
  \BibitemOpen
  \bibfield  {author} {\bibinfo {author} {\bibfnamefont {S.~L.}\ \bibnamefont
  {Braunstein}}, \bibinfo {author} {\bibfnamefont {C.~A.}\ \bibnamefont
  {Fuchs}}, \ and\ \bibinfo {author} {\bibfnamefont {H.~J.}\ \bibnamefont
  {Kimble}},\ }\href
  {http://www.tandfonline.com/doi/abs/10.1080/09500340008244041} {\bibfield
  {journal} {\bibinfo  {journal} {Journal of Modern Optics}\ }\textbf {\bibinfo
  {volume} {47}},\ \bibinfo {pages} {267} (\bibinfo {year} {2000})}\BibitemShut
  {NoStop}%
\bibitem [{\citenamefont {Bennett}\ and\ \citenamefont
  {Brassard}(1984)}]{BB84}%
  \BibitemOpen
  \bibfield  {author} {\bibinfo {author} {\bibfnamefont {C.~H.}\ \bibnamefont
  {Bennett}}\ and\ \bibinfo {author} {\bibfnamefont {G.}~\bibnamefont
  {Brassard}},\ }in\ \href@noop {} {\emph {\bibinfo {booktitle} {International
  Conference on Computer System and Signal Processing, IEEE, 1984}}}\ (\bibinfo
  {year} {1984})\ pp.\ \bibinfo {pages} {175--179}\BibitemShut {NoStop}%
\bibitem [{\citenamefont {Gisin}\ \emph {et~al.}(2002)\citenamefont {Gisin},
  \citenamefont {Ribordy}, \citenamefont {Tittel},\ and\ \citenamefont
  {Zbinden}}]{GRG+02}%
  \BibitemOpen
  \bibfield  {author} {\bibinfo {author} {\bibfnamefont {N.}~\bibnamefont
  {Gisin}}, \bibinfo {author} {\bibfnamefont {G.}~\bibnamefont {Ribordy}},
  \bibinfo {author} {\bibfnamefont {W.}~\bibnamefont {Tittel}}, \ and\ \bibinfo
  {author} {\bibfnamefont {H.}~\bibnamefont {Zbinden}},\ }\href {\doibase
  10.1103/RevModPhys.74.145} {\bibfield  {journal} {\bibinfo  {journal}
  {Reviews of Modern Physics}\ }\textbf {\bibinfo {volume} {74}},\ \bibinfo
  {pages} {145} (\bibinfo {year} {2002})}\BibitemShut {NoStop}%
\bibitem [{\citenamefont {Scarani}\ \emph {et~al.}(2009)\citenamefont
  {Scarani}, \citenamefont {Bechmann-Pasquinucci}, \citenamefont {Cerf},
  \citenamefont {Du\ifmmode~\check{s}\else \v{s}\fi{}ek}, \citenamefont
  {L\"{u}tkenhaus},\ and\ \citenamefont {Peev}}]{SBPC+09}%
  \BibitemOpen
  \bibfield  {author} {\bibinfo {author} {\bibfnamefont {V.}~\bibnamefont
  {Scarani}}, \bibinfo {author} {\bibfnamefont {H.}~\bibnamefont
  {Bechmann-Pasquinucci}}, \bibinfo {author} {\bibfnamefont {N.~J.}\
  \bibnamefont {Cerf}}, \bibinfo {author} {\bibfnamefont {M.}~\bibnamefont
  {Du\ifmmode~\check{s}\else \v{s}\fi{}ek}}, \bibinfo {author} {\bibfnamefont
  {N.}~\bibnamefont {L\"{u}tkenhaus}}, \ and\ \bibinfo {author} {\bibfnamefont
  {M.}~\bibnamefont {Peev}},\ }\href {\doibase 10.1103/RevModPhys.81.1301}
  {\bibfield  {journal} {\bibinfo  {journal} {Reviews of Modern Physics}\
  }\textbf {\bibinfo {volume} {81}},\ \bibinfo {pages} {1301} (\bibinfo {year}
  {2009})}\BibitemShut {NoStop}%
\bibitem [{\citenamefont {Cirac}\ \emph {et~al.}(1999)\citenamefont {Cirac},
  \citenamefont {Ekert}, \citenamefont {Huelga},\ and\ \citenamefont
  {Macchiavello}}]{CEHM99}%
  \BibitemOpen
  \bibfield  {author} {\bibinfo {author} {\bibfnamefont {J.~I.}\ \bibnamefont
  {Cirac}}, \bibinfo {author} {\bibfnamefont {A.~K.}\ \bibnamefont {Ekert}},
  \bibinfo {author} {\bibfnamefont {S.~F.}\ \bibnamefont {Huelga}}, \ and\
  \bibinfo {author} {\bibfnamefont {C.}~\bibnamefont {Macchiavello}},\ }\href
  {\doibase 10.1103/PhysRevA.59.4249} {\bibfield  {journal} {\bibinfo
  {journal} {Physical Review A}\ }\textbf {\bibinfo {volume} {59}},\ \bibinfo
  {pages} {4249} (\bibinfo {year} {1999})}\BibitemShut {NoStop}%
\bibitem [{\citenamefont {Aspect}\ \emph {et~al.}(1982)\citenamefont {Aspect},
  \citenamefont {Grangier},\ and\ \citenamefont {Roger}}]{AGR82}%
  \BibitemOpen
  \bibfield  {author} {\bibinfo {author} {\bibfnamefont {A.}~\bibnamefont
  {Aspect}}, \bibinfo {author} {\bibfnamefont {P.}~\bibnamefont {Grangier}}, \
  and\ \bibinfo {author} {\bibfnamefont {G.}~\bibnamefont {Roger}},\ }\href
  {\doibase 10.1103/PhysRevLett.49.91} {\bibfield  {journal} {\bibinfo
  {journal} {Physical Review Letters}\ }\textbf {\bibinfo {volume} {49}},\
  \bibinfo {pages} {91} (\bibinfo {year} {1982})}\BibitemShut {NoStop}%
\bibitem [{\citenamefont {Hensen}\ \emph {et~al.}(2015)\citenamefont {Hensen},
  \citenamefont {Bernien}, \citenamefont {Dr{\'e}au}, \citenamefont {Reiserer},
  \citenamefont {Kalb}, \citenamefont {Blok}, \citenamefont {Ruitenberg},
  \citenamefont {Vermeulen}, \citenamefont {Schouten}, \citenamefont
  {Abell{\'a}n} \emph {et~al.}}]{HBD+15}%
  \BibitemOpen
  \bibfield  {author} {\bibinfo {author} {\bibfnamefont {B.}~\bibnamefont
  {Hensen}}, \bibinfo {author} {\bibfnamefont {H.}~\bibnamefont {Bernien}},
  \bibinfo {author} {\bibfnamefont {A.~E.}\ \bibnamefont {Dr{\'e}au}}, \bibinfo
  {author} {\bibfnamefont {A.}~\bibnamefont {Reiserer}}, \bibinfo {author}
  {\bibfnamefont {N.}~\bibnamefont {Kalb}}, \bibinfo {author} {\bibfnamefont
  {M.~S.}\ \bibnamefont {Blok}}, \bibinfo {author} {\bibfnamefont
  {J.}~\bibnamefont {Ruitenberg}}, \bibinfo {author} {\bibfnamefont {R.~F.~L.}\
  \bibnamefont {Vermeulen}}, \bibinfo {author} {\bibfnamefont {R.~N.}\
  \bibnamefont {Schouten}}, \bibinfo {author} {\bibfnamefont {C.}~\bibnamefont
  {Abell{\'a}n}},  \emph {et~al.},\ }\href
  {http://dx.doi.org/10.1038/nature15759} {\bibfield  {journal} {\bibinfo
  {journal} {Nature}\ }\textbf {\bibinfo {volume} {526}},\ \bibinfo {pages}
  {682} (\bibinfo {year} {2015})}\BibitemShut {NoStop}%
\bibitem [{\citenamefont {Rosenfeld}\ \emph {et~al.}(2017)\citenamefont
  {Rosenfeld}, \citenamefont {Burchardt}, \citenamefont {Garthoff},
  \citenamefont {Redeker}, \citenamefont {Ortegel}, \citenamefont {Rau},\ and\
  \citenamefont {Weinfurter}}]{RBG17}%
  \BibitemOpen
  \bibfield  {author} {\bibinfo {author} {\bibfnamefont {W.}~\bibnamefont
  {Rosenfeld}}, \bibinfo {author} {\bibfnamefont {D.}~\bibnamefont
  {Burchardt}}, \bibinfo {author} {\bibfnamefont {R.}~\bibnamefont {Garthoff}},
  \bibinfo {author} {\bibfnamefont {K.}~\bibnamefont {Redeker}}, \bibinfo
  {author} {\bibfnamefont {N.}~\bibnamefont {Ortegel}}, \bibinfo {author}
  {\bibfnamefont {M.}~\bibnamefont {Rau}}, \ and\ \bibinfo {author}
  {\bibfnamefont {H.}~\bibnamefont {Weinfurter}},\ }\href {\doibase
  10.1103/PhysRevLett.119.010402} {\bibfield  {journal} {\bibinfo  {journal}
  {Physical Review Letters}\ }\textbf {\bibinfo {volume} {119}},\ \bibinfo
  {pages} {010402} (\bibinfo {year} {2017})}\BibitemShut {NoStop}%
\bibitem [{\citenamefont {Jozsa}\ \emph {et~al.}(2000)\citenamefont {Jozsa},
  \citenamefont {Abrams}, \citenamefont {Dowling},\ and\ \citenamefont
  {Williams}}]{JADW00}%
  \BibitemOpen
  \bibfield  {author} {\bibinfo {author} {\bibfnamefont {R.}~\bibnamefont
  {Jozsa}}, \bibinfo {author} {\bibfnamefont {D.~S.}\ \bibnamefont {Abrams}},
  \bibinfo {author} {\bibfnamefont {J.~P.}\ \bibnamefont {Dowling}}, \ and\
  \bibinfo {author} {\bibfnamefont {C.~P.}\ \bibnamefont {Williams}},\ }\href
  {\doibase 10.1103/PhysRevLett.85.2010} {\bibfield  {journal} {\bibinfo
  {journal} {Physical Review Letters}\ }\textbf {\bibinfo {volume} {85}},\
  \bibinfo {pages} {2010} (\bibinfo {year} {2000})}\BibitemShut {NoStop}%
\bibitem [{\citenamefont {Yurtsever}\ and\ \citenamefont
  {Dowling}(2002)}]{UD02}%
  \BibitemOpen
  \bibfield  {author} {\bibinfo {author} {\bibfnamefont {U.}~\bibnamefont
  {Yurtsever}}\ and\ \bibinfo {author} {\bibfnamefont {J.~P.}\ \bibnamefont
  {Dowling}},\ }\href {\doibase 10.1103/PhysRevA.65.052317} {\bibfield
  {journal} {\bibinfo  {journal} {Physical Review A}\ }\textbf {\bibinfo
  {volume} {65}},\ \bibinfo {pages} {052317} (\bibinfo {year}
  {2002})}\BibitemShut {NoStop}%
\bibitem [{\citenamefont {Ilo-Okeke}\ \emph {et~al.}(2017)\citenamefont
  {Ilo-Okeke}, \citenamefont {Tessler}, \citenamefont {Dowling},\ and\
  \citenamefont {Byrnes}}]{ITDB17}%
  \BibitemOpen
  \bibfield  {author} {\bibinfo {author} {\bibfnamefont {E.~O.}\ \bibnamefont
  {Ilo-Okeke}}, \bibinfo {author} {\bibfnamefont {L.}~\bibnamefont {Tessler}},
  \bibinfo {author} {\bibfnamefont {J.~P.}\ \bibnamefont {Dowling}}, \ and\
  \bibinfo {author} {\bibfnamefont {T.}~\bibnamefont {Byrnes}},\ }\href
  {https://arxiv.org/abs/1709.08423} {\bibfield  {journal} {\bibinfo  {journal}
  {arXiv:1709.08423}\ } (\bibinfo {year} {2017})}\BibitemShut {NoStop}%
\bibitem [{\citenamefont {Hillery}\ \emph {et~al.}(1999)\citenamefont
  {Hillery}, \citenamefont {Bu\ifmmode~\check{z}\else \v{z}\fi{}ek},\ and\
  \citenamefont {Berthiaume}}]{HBB99}%
  \BibitemOpen
  \bibfield  {author} {\bibinfo {author} {\bibfnamefont {M.}~\bibnamefont
  {Hillery}}, \bibinfo {author} {\bibfnamefont {V.}~\bibnamefont
  {Bu\ifmmode~\check{z}\else \v{z}\fi{}ek}}, \ and\ \bibinfo {author}
  {\bibfnamefont {A.}~\bibnamefont {Berthiaume}},\ }\href {\doibase
  10.1103/PhysRevA.59.1829} {\bibfield  {journal} {\bibinfo  {journal}
  {Physical Review A}\ }\textbf {\bibinfo {volume} {59}},\ \bibinfo {pages}
  {1829} (\bibinfo {year} {1999})}\BibitemShut {NoStop}%
\bibitem [{\citenamefont {Briegel}\ \emph {et~al.}(1998)\citenamefont
  {Briegel}, \citenamefont {D\"ur}, \citenamefont {Cirac},\ and\ \citenamefont
  {Zoller}}]{BDC98}%
  \BibitemOpen
  \bibfield  {author} {\bibinfo {author} {\bibfnamefont {H.-J.}\ \bibnamefont
  {Briegel}}, \bibinfo {author} {\bibfnamefont {W.}~\bibnamefont {D\"ur}},
  \bibinfo {author} {\bibfnamefont {J.~I.}\ \bibnamefont {Cirac}}, \ and\
  \bibinfo {author} {\bibfnamefont {P.}~\bibnamefont {Zoller}},\ }\href
  {\doibase 10.1103/PhysRevLett.81.5932} {\bibfield  {journal} {\bibinfo
  {journal} {Physical Review Letters}\ }\textbf {\bibinfo {volume} {81}},\
  \bibinfo {pages} {5932} (\bibinfo {year} {1998})}\BibitemShut {NoStop}%
\bibitem [{\citenamefont {D\"ur}\ \emph {et~al.}(1999)\citenamefont {D\"ur},
  \citenamefont {Briegel}, \citenamefont {Cirac},\ and\ \citenamefont
  {Zoller}}]{DBC99}%
  \BibitemOpen
  \bibfield  {author} {\bibinfo {author} {\bibfnamefont {W.}~\bibnamefont
  {D\"ur}}, \bibinfo {author} {\bibfnamefont {H.-J.}\ \bibnamefont {Briegel}},
  \bibinfo {author} {\bibfnamefont {J.~I.}\ \bibnamefont {Cirac}}, \ and\
  \bibinfo {author} {\bibfnamefont {P.}~\bibnamefont {Zoller}},\ }\href
  {\doibase 10.1103/PhysRevA.59.169} {\bibfield  {journal} {\bibinfo  {journal}
  {Physical Review A}\ }\textbf {\bibinfo {volume} {59}},\ \bibinfo {pages}
  {169} (\bibinfo {year} {1999})}\BibitemShut {NoStop}%
\bibitem [{\citenamefont {\ifmmode~\dot{Z}\else \.{Z}\fi{}ukowski}\ \emph
  {et~al.}(1993)\citenamefont {\ifmmode~\dot{Z}\else \.{Z}\fi{}ukowski},
  \citenamefont {Zeilinger}, \citenamefont {Horne},\ and\ \citenamefont
  {Ekert}}]{ZZH93}%
  \BibitemOpen
  \bibfield  {author} {\bibinfo {author} {\bibfnamefont {M.}~\bibnamefont
  {\ifmmode~\dot{Z}\else \.{Z}\fi{}ukowski}}, \bibinfo {author} {\bibfnamefont
  {A.}~\bibnamefont {Zeilinger}}, \bibinfo {author} {\bibfnamefont {M.~A.}\
  \bibnamefont {Horne}}, \ and\ \bibinfo {author} {\bibfnamefont {A.~K.}\
  \bibnamefont {Ekert}},\ }\href {\doibase 10.1103/PhysRevLett.71.4287}
  {\bibfield  {journal} {\bibinfo  {journal} {Physical Review Letters}\
  }\textbf {\bibinfo {volume} {71}},\ \bibinfo {pages} {4287} (\bibinfo {year}
  {1993})}\BibitemShut {NoStop}%
\bibitem [{\citenamefont {Bennett}\ \emph
  {et~al.}(1996{\natexlab{a}})\citenamefont {Bennett}, \citenamefont
  {Brassard}, \citenamefont {Popescu}, \citenamefont {Schumacher},
  \citenamefont {Smolin},\ and\ \citenamefont {Wootters}}]{BBP96}%
  \BibitemOpen
  \bibfield  {author} {\bibinfo {author} {\bibfnamefont {C.~H.}\ \bibnamefont
  {Bennett}}, \bibinfo {author} {\bibfnamefont {G.}~\bibnamefont {Brassard}},
  \bibinfo {author} {\bibfnamefont {S.}~\bibnamefont {Popescu}}, \bibinfo
  {author} {\bibfnamefont {B.}~\bibnamefont {Schumacher}}, \bibinfo {author}
  {\bibfnamefont {J.~A.}\ \bibnamefont {Smolin}}, \ and\ \bibinfo {author}
  {\bibfnamefont {W.~K.}\ \bibnamefont {Wootters}},\ }\href {\doibase
  10.1103/PhysRevLett.76.722} {\bibfield  {journal} {\bibinfo  {journal}
  {Physical Review Letters}\ }\textbf {\bibinfo {volume} {76}},\ \bibinfo
  {pages} {722} (\bibinfo {year} {1996}{\natexlab{a}})}\BibitemShut {NoStop}%
\bibitem [{\citenamefont {Deutsch}\ \emph {et~al.}(1996)\citenamefont
  {Deutsch}, \citenamefont {Ekert}, \citenamefont {Jozsa}, \citenamefont
  {Macchiavello}, \citenamefont {Popescu},\ and\ \citenamefont
  {Sanpera}}]{DAR96}%
  \BibitemOpen
  \bibfield  {author} {\bibinfo {author} {\bibfnamefont {D.}~\bibnamefont
  {Deutsch}}, \bibinfo {author} {\bibfnamefont {A.}~\bibnamefont {Ekert}},
  \bibinfo {author} {\bibfnamefont {R.}~\bibnamefont {Jozsa}}, \bibinfo
  {author} {\bibfnamefont {C.}~\bibnamefont {Macchiavello}}, \bibinfo {author}
  {\bibfnamefont {S.}~\bibnamefont {Popescu}}, \ and\ \bibinfo {author}
  {\bibfnamefont {A.}~\bibnamefont {Sanpera}},\ }\href {\doibase
  10.1103/PhysRevLett.77.2818} {\bibfield  {journal} {\bibinfo  {journal}
  {Physical Review Letters}\ }\textbf {\bibinfo {volume} {77}},\ \bibinfo
  {pages} {2818} (\bibinfo {year} {1996})}\BibitemShut {NoStop}%
\bibitem [{\citenamefont {Bennett}\ \emph
  {et~al.}(1996{\natexlab{b}})\citenamefont {Bennett}, \citenamefont
  {DiVincenzo}, \citenamefont {Smolin},\ and\ \citenamefont
  {Wootters}}]{BDD96}%
  \BibitemOpen
  \bibfield  {author} {\bibinfo {author} {\bibfnamefont {C.~H.}\ \bibnamefont
  {Bennett}}, \bibinfo {author} {\bibfnamefont {D.~P.}\ \bibnamefont
  {DiVincenzo}}, \bibinfo {author} {\bibfnamefont {J.~A.}\ \bibnamefont
  {Smolin}}, \ and\ \bibinfo {author} {\bibfnamefont {W.~K.}\ \bibnamefont
  {Wootters}},\ }\href {\doibase 10.1103/PhysRevA.54.3824} {\bibfield
  {journal} {\bibinfo  {journal} {Physical Review A}\ }\textbf {\bibinfo
  {volume} {54}},\ \bibinfo {pages} {3824} (\bibinfo {year}
  {1996}{\natexlab{b}})}\BibitemShut {NoStop}%
\bibitem [{\citenamefont {Azuma}\ \emph {et~al.}(2015)\citenamefont {Azuma},
  \citenamefont {Tamaki},\ and\ \citenamefont {Lo}}]{KKL15}%
  \BibitemOpen
  \bibfield  {author} {\bibinfo {author} {\bibfnamefont {K.}~\bibnamefont
  {Azuma}}, \bibinfo {author} {\bibfnamefont {K.}~\bibnamefont {Tamaki}}, \
  and\ \bibinfo {author} {\bibfnamefont {H.-K.}\ \bibnamefont {Lo}},\ }\href
  {http://dx.doi.org/10.1038/ncomms7787} {\bibfield  {journal} {\bibinfo
  {journal} {Nature Communications}\ }\textbf {\bibinfo {volume} {6}} (\bibinfo
  {year} {2015})}\BibitemShut {NoStop}%
\bibitem [{\citenamefont {Miatto}\ \emph {et~al.}(2017)\citenamefont {Miatto},
  \citenamefont {Epping},\ and\ \citenamefont {L\"{u}tkenhaus}}]{MEL17}%
  \BibitemOpen
  \bibfield  {author} {\bibinfo {author} {\bibfnamefont {F.~M.}\ \bibnamefont
  {Miatto}}, \bibinfo {author} {\bibfnamefont {M.}~\bibnamefont {Epping}}, \
  and\ \bibinfo {author} {\bibfnamefont {N.}~\bibnamefont {L\"{u}tkenhaus}},\
  }\href {https://arxiv.org/abs/1710.03063} {\bibfield  {journal} {\bibinfo
  {journal} {arXiv:1710.03063}\ } (\bibinfo {year} {2017})}\BibitemShut
  {NoStop}%
\bibitem [{\citenamefont {Takeoka}\ \emph {et~al.}(2015)\citenamefont
  {Takeoka}, \citenamefont {Guha},\ and\ \citenamefont {Wilde}}]{TGW15}%
  \BibitemOpen
  \bibfield  {author} {\bibinfo {author} {\bibfnamefont {M.}~\bibnamefont
  {Takeoka}}, \bibinfo {author} {\bibfnamefont {S.}~\bibnamefont {Guha}}, \
  and\ \bibinfo {author} {\bibfnamefont {M.~M.}\ \bibnamefont {Wilde}},\ }\href
  {http://dx.doi.org/10.1038/ncomms6235} {\bibfield  {journal} {\bibinfo
  {journal} {Nature Communications}\ }\textbf {\bibinfo {volume} {5}} (\bibinfo
  {year} {2015})}\BibitemShut {NoStop}%
\bibitem [{\citenamefont {Pirandola}\ \emph {et~al.}(2017)\citenamefont
  {Pirandola}, \citenamefont {Laurenza}, \citenamefont {Ottaviani},\ and\
  \citenamefont {Banchi}}]{PLOB17}%
  \BibitemOpen
  \bibfield  {author} {\bibinfo {author} {\bibfnamefont {S.}~\bibnamefont
  {Pirandola}}, \bibinfo {author} {\bibfnamefont {R.}~\bibnamefont {Laurenza}},
  \bibinfo {author} {\bibfnamefont {C.}~\bibnamefont {Ottaviani}}, \ and\
  \bibinfo {author} {\bibfnamefont {L.}~\bibnamefont {Banchi}},\ }\href
  {http://dx.doi.org/10.1038/ncomms15043} {\bibfield  {journal} {\bibinfo
  {journal} {Nature Communications}\ }\textbf {\bibinfo {volume} {8}} (\bibinfo
  {year} {2017})}\BibitemShut {NoStop}%
\bibitem [{\citenamefont {Garc\'{i}a-Patr\'{o}n}\ \emph
  {et~al.}(2009)\citenamefont {Garc\'{i}a-Patr\'{o}n}, \citenamefont
  {Pirandola}, \citenamefont {Lloyd},\ and\ \citenamefont {Shapiro}}]{GPLS09}%
  \BibitemOpen
  \bibfield  {author} {\bibinfo {author} {\bibfnamefont {R.}~\bibnamefont
  {Garc\'{i}a-Patr\'{o}n}}, \bibinfo {author} {\bibfnamefont {S.}~\bibnamefont
  {Pirandola}}, \bibinfo {author} {\bibfnamefont {S.}~\bibnamefont {Lloyd}}, \
  and\ \bibinfo {author} {\bibfnamefont {J.~H.}\ \bibnamefont {Shapiro}},\
  }\href {\doibase 10.1103/PhysRevLett.102.210501} {\bibfield  {journal}
  {\bibinfo  {journal} {Physical Review Letters}\ }\textbf {\bibinfo {volume}
  {102}},\ \bibinfo {pages} {210501} (\bibinfo {year} {2009})}\BibitemShut
  {NoStop}%
\bibitem [{\citenamefont {Wilde}\ \emph {et~al.}(2017)\citenamefont {Wilde},
  \citenamefont {Tomamichel},\ and\ \citenamefont {Berta}}]{WTB17}%
  \BibitemOpen
  \bibfield  {author} {\bibinfo {author} {\bibfnamefont {M.~M.}\ \bibnamefont
  {Wilde}}, \bibinfo {author} {\bibfnamefont {M.}~\bibnamefont {Tomamichel}}, \
  and\ \bibinfo {author} {\bibfnamefont {M.}~\bibnamefont {Berta}},\ }\href
  {\doibase 10.1109/TIT.2017.2648825} {\bibfield  {journal} {\bibinfo
  {journal} {IEEE Transactions on Information Theory}\ }\textbf {\bibinfo
  {volume} {63}},\ \bibinfo {pages} {1792} (\bibinfo {year}
  {2017})}\BibitemShut {NoStop}%
\bibitem [{\citenamefont {Ralph}\ and\ \citenamefont {Pryde}(2010)}]{RP10}%
  \BibitemOpen
  \bibfield  {author} {\bibinfo {author} {\bibfnamefont {T.~C.}\ \bibnamefont
  {Ralph}}\ and\ \bibinfo {author} {\bibfnamefont {G.~J.}\ \bibnamefont
  {Pryde}},\ }\enquote {\bibinfo {title} {Optical quantum computation},}\ \
  (\bibinfo  {publisher} {Elsevier},\ \bibinfo {year} {2010})\ Chap.~\bibinfo
  {chapter} {4}, pp.\ \bibinfo {pages} {209--269}\BibitemShut {NoStop}%
\bibitem [{\citenamefont {Jacobs}\ \emph {et~al.}(2002)\citenamefont {Jacobs},
  \citenamefont {Pittman},\ and\ \citenamefont {Franson}}]{JPF02}%
  \BibitemOpen
  \bibfield  {author} {\bibinfo {author} {\bibfnamefont {B.~C.}\ \bibnamefont
  {Jacobs}}, \bibinfo {author} {\bibfnamefont {T.~B.}\ \bibnamefont {Pittman}},
  \ and\ \bibinfo {author} {\bibfnamefont {J.~D.}\ \bibnamefont {Franson}},\
  }\href {\doibase 10.1103/PhysRevA.66.052307} {\bibfield  {journal} {\bibinfo
  {journal} {Physical Review A}\ }\textbf {\bibinfo {volume} {66}},\ \bibinfo
  {pages} {052307} (\bibinfo {year} {2002})}\BibitemShut {NoStop}%
\bibitem [{\citenamefont {de~Riedmatten}\ \emph {et~al.}(2004)\citenamefont
  {de~Riedmatten}, \citenamefont {Marcikic}, \citenamefont {Tittel},
  \citenamefont {Zbinden}, \citenamefont {Collins},\ and\ \citenamefont
  {Gisin}}]{DMT+04}%
  \BibitemOpen
  \bibfield  {author} {\bibinfo {author} {\bibfnamefont {H.}~\bibnamefont
  {de~Riedmatten}}, \bibinfo {author} {\bibfnamefont {I.}~\bibnamefont
  {Marcikic}}, \bibinfo {author} {\bibfnamefont {W.}~\bibnamefont {Tittel}},
  \bibinfo {author} {\bibfnamefont {H.}~\bibnamefont {Zbinden}}, \bibinfo
  {author} {\bibfnamefont {D.}~\bibnamefont {Collins}}, \ and\ \bibinfo
  {author} {\bibfnamefont {N.}~\bibnamefont {Gisin}},\ }\href {\doibase
  10.1103/PhysRevLett.92.047904} {\bibfield  {journal} {\bibinfo  {journal}
  {Physical Review Letters}\ }\textbf {\bibinfo {volume} {92}},\ \bibinfo
  {pages} {047904} (\bibinfo {year} {2004})}\BibitemShut {NoStop}%
\bibitem [{\citenamefont {Collins}\ \emph {et~al.}(2005)\citenamefont
  {Collins}, \citenamefont {Gisin},\ and\ \citenamefont {Riedmatten}}]{CGD05}%
  \BibitemOpen
  \bibfield  {author} {\bibinfo {author} {\bibfnamefont {D.}~\bibnamefont
  {Collins}}, \bibinfo {author} {\bibfnamefont {N.}~\bibnamefont {Gisin}}, \
  and\ \bibinfo {author} {\bibfnamefont {H.~D.}\ \bibnamefont {Riedmatten}},\
  }\href {\doibase 10.1080/09500340412331283633} {\bibfield  {journal}
  {\bibinfo  {journal} {Journal of Modern Optics}\ }\textbf {\bibinfo {volume}
  {52}},\ \bibinfo {pages} {735} (\bibinfo {year} {2005})}\BibitemShut
  {NoStop}%
\bibitem [{\citenamefont {Sangouard}\ \emph {et~al.}(2011)\citenamefont
  {Sangouard}, \citenamefont {Simon}, \citenamefont {de~Riedmatten},\ and\
  \citenamefont {Gisin}}]{SSR+11}%
  \BibitemOpen
  \bibfield  {author} {\bibinfo {author} {\bibfnamefont {N.}~\bibnamefont
  {Sangouard}}, \bibinfo {author} {\bibfnamefont {C.}~\bibnamefont {Simon}},
  \bibinfo {author} {\bibfnamefont {H.}~\bibnamefont {de~Riedmatten}}, \ and\
  \bibinfo {author} {\bibfnamefont {N.}~\bibnamefont {Gisin}},\ }\href
  {\doibase 10.1103/RevModPhys.83.33} {\bibfield  {journal} {\bibinfo
  {journal} {Reviews of Modern Physics}\ }\textbf {\bibinfo {volume} {83}},\
  \bibinfo {pages} {33} (\bibinfo {year} {2011})}\BibitemShut {NoStop}%
\bibitem [{\citenamefont {Mayers}\ and\ \citenamefont {Yao}(1998)}]{MY98}%
  \BibitemOpen
  \bibfield  {author} {\bibinfo {author} {\bibfnamefont {D.}~\bibnamefont
  {Mayers}}\ and\ \bibinfo {author} {\bibfnamefont {A.}~\bibnamefont {Yao}},\
  }in\ \href {http://dl.acm.org/citation.cfm?id=795664.796390} {\emph {\bibinfo
  {booktitle} {Proceedings of the 39th Annual Symposium on Foundations of
  Computer Science}}},\ \bibinfo {series and number} {FOCS '98}\ (\bibinfo
  {publisher} {IEEE Computer Society},\ \bibinfo {address} {Washington, DC,
  USA},\ \bibinfo {year} {1998})\ pp.\ \bibinfo {pages} {503--509}\BibitemShut
  {NoStop}%
\bibitem [{\citenamefont {Ac\'{i}n}\ \emph {et~al.}(2007)\citenamefont
  {Ac\'{i}n}, \citenamefont {Brunner}, \citenamefont {Gisin}, \citenamefont
  {Massar}, \citenamefont {Pironio},\ and\ \citenamefont {Scarani}}]{ABG+07}%
  \BibitemOpen
  \bibfield  {author} {\bibinfo {author} {\bibfnamefont {A.}~\bibnamefont
  {Ac\'{i}n}}, \bibinfo {author} {\bibfnamefont {N.}~\bibnamefont {Brunner}},
  \bibinfo {author} {\bibfnamefont {N.}~\bibnamefont {Gisin}}, \bibinfo
  {author} {\bibfnamefont {S.}~\bibnamefont {Massar}}, \bibinfo {author}
  {\bibfnamefont {S.}~\bibnamefont {Pironio}}, \ and\ \bibinfo {author}
  {\bibfnamefont {V.}~\bibnamefont {Scarani}},\ }\href {\doibase
  10.1103/PhysRevLett.98.230501} {\bibfield  {journal} {\bibinfo  {journal}
  {Physical Review Letters}\ }\textbf {\bibinfo {volume} {98}},\ \bibinfo
  {pages} {230501} (\bibinfo {year} {2007})}\BibitemShut {NoStop}%
\bibitem [{\citenamefont {Lo}\ \emph {et~al.}(2012)\citenamefont {Lo},
  \citenamefont {Curty},\ and\ \citenamefont {Qi}}]{LCQ12}%
  \BibitemOpen
  \bibfield  {author} {\bibinfo {author} {\bibfnamefont {H.-K.}\ \bibnamefont
  {Lo}}, \bibinfo {author} {\bibfnamefont {M.}~\bibnamefont {Curty}}, \ and\
  \bibinfo {author} {\bibfnamefont {B.}~\bibnamefont {Qi}},\ }\href {\doibase
  10.1103/PhysRevLett.108.130503} {\bibfield  {journal} {\bibinfo  {journal}
  {Physical Review Letters}\ }\textbf {\bibinfo {volume} {108}},\ \bibinfo
  {pages} {130503} (\bibinfo {year} {2012})}\BibitemShut {NoStop}%
\bibitem [{\citenamefont {Braunstein}\ and\ \citenamefont
  {Pirandola}(2012)}]{BP12}%
  \BibitemOpen
  \bibfield  {author} {\bibinfo {author} {\bibfnamefont {S.~L.}\ \bibnamefont
  {Braunstein}}\ and\ \bibinfo {author} {\bibfnamefont {S.}~\bibnamefont
  {Pirandola}},\ }\href {\doibase 10.1103/PhysRevLett.108.130502} {\bibfield
  {journal} {\bibinfo  {journal} {Physical Review Letters}\ }\textbf {\bibinfo
  {volume} {108}},\ \bibinfo {pages} {130502} (\bibinfo {year}
  {2012})}\BibitemShut {NoStop}%
\bibitem [{\citenamefont {Bognat}\ and\ \citenamefont {Hayden}(2014)}]{BH14}%
  \BibitemOpen
  \bibfield  {author} {\bibinfo {author} {\bibfnamefont {A.}~\bibnamefont
  {Bognat}}\ and\ \bibinfo {author} {\bibfnamefont {P.}~\bibnamefont
  {Hayden}},\ }\enquote {\bibinfo {title} {Privacy from accelerating
  eavesdroppers: The impact of losses},}\ in\ \href {\doibase
  10.1007/978-3-319-06880-0_9} {\emph {\bibinfo {booktitle} {Horizons of the
  Mind. A Tribute to Prakash Panangaden: Essays Dedicated to Prakash Panangaden
  on the Occasion of His 60th Birthday}}},\ \bibinfo {editor} {edited by\
  \bibinfo {editor} {\bibfnamefont {F.}~\bibnamefont {van Breugel}}, \bibinfo
  {editor} {\bibfnamefont {E.}~\bibnamefont {Kashefi}}, \bibinfo {editor}
  {\bibfnamefont {C.}~\bibnamefont {Palamidessi}}, \ and\ \bibinfo {editor}
  {\bibfnamefont {J.}~\bibnamefont {Rutten}}}\ (\bibinfo  {publisher} {Springer
  International Publishing},\ \bibinfo {address} {Cham},\ \bibinfo {year}
  {2014})\ pp.\ \bibinfo {pages} {180--190}\BibitemShut {NoStop}%
\bibitem [{\citenamefont {Grassl}\ \emph {et~al.}(1997)\citenamefont {Grassl},
  \citenamefont {Beth},\ and\ \citenamefont {Pellizzari}}]{GBP97}%
  \BibitemOpen
  \bibfield  {author} {\bibinfo {author} {\bibfnamefont {M.}~\bibnamefont
  {Grassl}}, \bibinfo {author} {\bibfnamefont {T.}~\bibnamefont {Beth}}, \ and\
  \bibinfo {author} {\bibfnamefont {T.}~\bibnamefont {Pellizzari}},\ }\href
  {\doibase 10.1103/PhysRevA.56.33} {\bibfield  {journal} {\bibinfo  {journal}
  {Physical Review A}\ }\textbf {\bibinfo {volume} {56}},\ \bibinfo {pages}
  {33} (\bibinfo {year} {1997})}\BibitemShut {NoStop}%
\bibitem [{\citenamefont {Vaidman}\ and\ \citenamefont {Yoran}(1999)}]{VY99}%
  \BibitemOpen
  \bibfield  {author} {\bibinfo {author} {\bibfnamefont {L.}~\bibnamefont
  {Vaidman}}\ and\ \bibinfo {author} {\bibfnamefont {N.}~\bibnamefont
  {Yoran}},\ }\href {\doibase 10.1103/PhysRevA.59.116} {\bibfield  {journal}
  {\bibinfo  {journal} {Physical Review A}\ }\textbf {\bibinfo {volume} {59}},\
  \bibinfo {pages} {116} (\bibinfo {year} {1999})}\BibitemShut {NoStop}%
\bibitem [{\citenamefont {L\"utkenhaus}\ \emph {et~al.}(1999)\citenamefont
  {L\"utkenhaus}, \citenamefont {Calsamiglia},\ and\ \citenamefont
  {Suominen}}]{LCS99}%
  \BibitemOpen
  \bibfield  {author} {\bibinfo {author} {\bibfnamefont {N.}~\bibnamefont
  {L\"utkenhaus}}, \bibinfo {author} {\bibfnamefont {J.}~\bibnamefont
  {Calsamiglia}}, \ and\ \bibinfo {author} {\bibfnamefont {K.-A.}\ \bibnamefont
  {Suominen}},\ }\href {\doibase 10.1103/PhysRevA.59.3295} {\bibfield
  {journal} {\bibinfo  {journal} {Physical Review A}\ }\textbf {\bibinfo
  {volume} {59}},\ \bibinfo {pages} {3295} (\bibinfo {year}
  {1999})}\BibitemShut {NoStop}%
\bibitem [{\citenamefont {Calsamiglia}\ and\ \citenamefont
  {L{\"u}tkenhaus}(2001)}]{CL01}%
  \BibitemOpen
  \bibfield  {author} {\bibinfo {author} {\bibfnamefont {J.}~\bibnamefont
  {Calsamiglia}}\ and\ \bibinfo {author} {\bibfnamefont {N.}~\bibnamefont
  {L{\"u}tkenhaus}},\ }\href {\doibase 10.1007/s003400000484} {\bibfield
  {journal} {\bibinfo  {journal} {Applied Physics B}\ }\textbf {\bibinfo
  {volume} {72}},\ \bibinfo {pages} {67} (\bibinfo {year} {2001})}\BibitemShut
  {NoStop}%
\bibitem [{\citenamefont {Kim}\ \emph {et~al.}(2001)\citenamefont {Kim},
  \citenamefont {Kulik},\ and\ \citenamefont {Shih}}]{KKS01}%
  \BibitemOpen
  \bibfield  {author} {\bibinfo {author} {\bibfnamefont {Y.-H.}\ \bibnamefont
  {Kim}}, \bibinfo {author} {\bibfnamefont {S.~P.}\ \bibnamefont {Kulik}}, \
  and\ \bibinfo {author} {\bibfnamefont {Y.}~\bibnamefont {Shih}},\ }\href
  {\doibase 10.1103/PhysRevLett.86.1370} {\bibfield  {journal} {\bibinfo
  {journal} {Physical Review Letters}\ }\textbf {\bibinfo {volume} {86}},\
  \bibinfo {pages} {1370} (\bibinfo {year} {2001})}\BibitemShut {NoStop}%
\bibitem [{\citenamefont {Kim}\ \emph {et~al.}(2002)\citenamefont {Kim},
  \citenamefont {Kulik},\ and\ \citenamefont {Shih}}]{KKS02}%
  \BibitemOpen
  \bibfield  {author} {\bibinfo {author} {\bibfnamefont {Y.-H.}\ \bibnamefont
  {Kim}}, \bibinfo {author} {\bibfnamefont {S.}~\bibnamefont {Kulik}}, \ and\
  \bibinfo {author} {\bibfnamefont {Y.}~\bibnamefont {Shih}},\ }\href {\doibase
  10.1080/09500340110087633} {\bibfield  {journal} {\bibinfo  {journal}
  {Journal of Modern Optics}\ }\textbf {\bibinfo {volume} {49}},\ \bibinfo
  {pages} {221} (\bibinfo {year} {2002})},\ \Eprint
  {http://arxiv.org/abs/http://dx.doi.org/10.1080/09500340110087633}
  {http://dx.doi.org/10.1080/09500340110087633} \BibitemShut {NoStop}%
\bibitem [{\citenamefont {Schoute}\ \emph {et~al.}(2016)\citenamefont
  {Schoute}, \citenamefont {Mancinska}, \citenamefont {Islam}, \citenamefont
  {Kerenidis},\ and\ \citenamefont {Wehner}}]{SMI+16}%
  \BibitemOpen
  \bibfield  {author} {\bibinfo {author} {\bibfnamefont {E.}~\bibnamefont
  {Schoute}}, \bibinfo {author} {\bibfnamefont {L.}~\bibnamefont {Mancinska}},
  \bibinfo {author} {\bibfnamefont {T.}~\bibnamefont {Islam}}, \bibinfo
  {author} {\bibfnamefont {I.}~\bibnamefont {Kerenidis}}, \ and\ \bibinfo
  {author} {\bibfnamefont {S.}~\bibnamefont {Wehner}},\ }\href
  {http://arxiv.org/abs/1610.05238} {\bibfield  {journal} {\bibinfo  {journal}
  {arXiv:1610.05238}\ } (\bibinfo {year} {2016})}\BibitemShut {NoStop}%
\bibitem [{Note1()}]{Note1}%
  \BibitemOpen
  \bibinfo {note} {It should be noted that one could allow for a different
  network setup where each workstation is capable of generating entangled pair
  of photons when needed and entangled photons be stored in its quantum memory
  or transmitted to the neighboring workstations depending on the
  task.}\BibitemShut {Stop}%
\bibitem [{\citenamefont {Grimmett}(1999)}]{Gri99book}%
  \BibitemOpen
  \bibfield  {author} {\bibinfo {author} {\bibfnamefont {G.~R.}\ \bibnamefont
  {Grimmett}},\ }\href {\doibase 10.1007/978-3-662-03981-6} {\emph {\bibinfo
  {title} {Percolation}}},\ \bibinfo {edition} {2nd}\ ed.\ (\bibinfo
  {publisher} {Springer-Verlag Berlin Heidelberg},\ \bibinfo {year}
  {1999})\BibitemShut {NoStop}%
\bibitem [{Note2()}]{Note2}%
  \BibitemOpen
  \bibinfo {note} {Previous studies of entanglement distribution in large
  networks based on percolation theory can be found in Refs. \cite
  {ACL07,PCA+08,CC09,LWL09}.}\BibitemShut {Stop}%
\bibitem [{\citenamefont {Suding}\ and\ \citenamefont {Ziff}(1999)}]{SZ99}%
  \BibitemOpen
  \bibfield  {author} {\bibinfo {author} {\bibfnamefont {P.~N.}\ \bibnamefont
  {Suding}}\ and\ \bibinfo {author} {\bibfnamefont {R.~M.}\ \bibnamefont
  {Ziff}},\ }\href {\doibase 10.1103/PhysRevE.60.275} {\bibfield  {journal}
  {\bibinfo  {journal} {Physical Review E}\ }\textbf {\bibinfo {volume} {60}},\
  \bibinfo {pages} {275} (\bibinfo {year} {1999})}\BibitemShut {NoStop}%
\bibitem [{\citenamefont {Newman}\ and\ \citenamefont {Ziff}(2001)}]{NZ01}%
  \BibitemOpen
  \bibfield  {author} {\bibinfo {author} {\bibfnamefont {M.~E.~J.}\
  \bibnamefont {Newman}}\ and\ \bibinfo {author} {\bibfnamefont {R.~M.}\
  \bibnamefont {Ziff}},\ }\href {\doibase 10.1103/PhysRevE.64.016706}
  {\bibfield  {journal} {\bibinfo  {journal} {Physical Review E}\ }\textbf
  {\bibinfo {volume} {64}},\ \bibinfo {pages} {016706} (\bibinfo {year}
  {2001})}\BibitemShut {NoStop}%
\bibitem [{\citenamefont {Wierman}(1984)}]{W84}%
  \BibitemOpen
  \bibfield  {author} {\bibinfo {author} {\bibfnamefont {J.~C.}\ \bibnamefont
  {Wierman}},\ }\href {http://stacks.iop.org/0305-4470/17/i=7/a=020} {\bibfield
   {journal} {\bibinfo  {journal} {Journal of Physics A: Mathematical and
  General}\ }\textbf {\bibinfo {volume} {17}},\ \bibinfo {pages} {1525}
  (\bibinfo {year} {1984})}\BibitemShut {NoStop}%
\bibitem [{\citenamefont {Ziff}\ and\ \citenamefont {Scullard}(2006)}]{ZS06}%
  \BibitemOpen
  \bibfield  {author} {\bibinfo {author} {\bibfnamefont {R.~M.}\ \bibnamefont
  {Ziff}}\ and\ \bibinfo {author} {\bibfnamefont {C.~R.}\ \bibnamefont
  {Scullard}},\ }\href {http://stacks.iop.org/0305-4470/39/i=49/a=003}
  {\bibfield  {journal} {\bibinfo  {journal} {Journal of Physics A:
  Mathematical and General}\ }\textbf {\bibinfo {volume} {39}},\ \bibinfo
  {pages} {15083} (\bibinfo {year} {2006})}\BibitemShut {NoStop}%
\bibitem [{Note3()}]{Note3}%
  \BibitemOpen
  \bibinfo {note} {More generally, for any noise model in a homogeneous network
  topology, if $\beta $ is the probability of sharing perfect entanglement,
  i.e., a Bell pair, between any two neighboring nodes, then the critical
  probability $\beta _c$ of establishing a large cluster of entangled nodes is
  equal to $p_c^{\protect \text {bond}}$. Note that subsequent entanglement
  swapping operations to establish entanglement between two distant nodes in
  such a cluster may have to be supplemented with purification
  protocols.}\BibitemShut {Stop}%
\bibitem [{\citenamefont {van~der Marck}(1997)}]{VDM97}%
  \BibitemOpen
  \bibfield  {author} {\bibinfo {author} {\bibfnamefont {S.~C.}\ \bibnamefont
  {van~der Marck}},\ }\href {\doibase 10.1103/PhysRevE.55.1514} {\bibfield
  {journal} {\bibinfo  {journal} {Physical Review E}\ }\textbf {\bibinfo
  {volume} {55}},\ \bibinfo {pages} {1514} (\bibinfo {year}
  {1997})}\BibitemShut {NoStop}%
\bibitem [{\citenamefont {Parviainen}(2007)}]{P07}%
  \BibitemOpen
  \bibfield  {author} {\bibinfo {author} {\bibfnamefont {R.}~\bibnamefont
  {Parviainen}},\ }\href {http://stacks.iop.org/1751-8121/40/i=31/a=005}
  {\bibfield  {journal} {\bibinfo  {journal} {Journal of Physics A:
  Mathematical and Theoretical}\ }\textbf {\bibinfo {volume} {40}},\ \bibinfo
  {pages} {9253} (\bibinfo {year} {2007})}\BibitemShut {NoStop}%
\bibitem [{\citenamefont {Sykes}\ and\ \citenamefont {Essam}(1964)}]{SE64}%
  \BibitemOpen
  \bibfield  {author} {\bibinfo {author} {\bibfnamefont {M.~F.}\ \bibnamefont
  {Sykes}}\ and\ \bibinfo {author} {\bibfnamefont {J.~W.}\ \bibnamefont
  {Essam}},\ }\href {http://dx.doi.org/10.1063/1.1704215} {\bibfield  {journal}
  {\bibinfo  {journal} {Journal of Mathematical Physics}\ }\textbf {\bibinfo
  {volume} {5}},\ \bibinfo {pages} {1117} (\bibinfo {year} {1964})}\BibitemShut
  {NoStop}%
\bibitem [{\citenamefont {Jacobsen}(2014)}]{Jac14}%
  \BibitemOpen
  \bibfield  {author} {\bibinfo {author} {\bibfnamefont {J.~L.}\ \bibnamefont
  {Jacobsen}},\ }\href {http://stacks.iop.org/1751-8121/47/i=13/a=135001}
  {\bibfield  {journal} {\bibinfo  {journal} {Journal of Physics A:
  Mathematical and Theoretical}\ }\textbf {\bibinfo {volume} {47}},\ \bibinfo
  {pages} {135001} (\bibinfo {year} {2014})}\BibitemShut {NoStop}%
\bibitem [{\citenamefont {Van~Meter}(2014)}]{VanMeter_book}%
  \BibitemOpen
  \bibfield  {author} {\bibinfo {author} {\bibfnamefont {R.}~\bibnamefont
  {Van~Meter}},\ }\href {\doibase 10.1002/9781118648919} {\emph {\bibinfo
  {title} {Quantum Networking}}}\ (\bibinfo  {publisher} {John Wiley \& Sons,
  Ltd},\ \bibinfo {year} {2014})\BibitemShut {NoStop}%
\bibitem [{\citenamefont {Scullard}\ and\ \citenamefont {Ziff}(2006)}]{SZ06}%
  \BibitemOpen
  \bibfield  {author} {\bibinfo {author} {\bibfnamefont {C.~R.}\ \bibnamefont
  {Scullard}}\ and\ \bibinfo {author} {\bibfnamefont {R.~M.}\ \bibnamefont
  {Ziff}},\ }\href {\doibase 10.1103/PhysRevE.73.045102} {\bibfield  {journal}
  {\bibinfo  {journal} {Physical Review E}\ }\textbf {\bibinfo {volume} {73}},\
  \bibinfo {pages} {045102} (\bibinfo {year} {2006})}\BibitemShut {NoStop}%
\bibitem [{\citenamefont {Scullard}\ and\ \citenamefont {Ziff}(2008)}]{SZ08}%
  \BibitemOpen
  \bibfield  {author} {\bibinfo {author} {\bibfnamefont {C.~R.}\ \bibnamefont
  {Scullard}}\ and\ \bibinfo {author} {\bibfnamefont {R.~M.}\ \bibnamefont
  {Ziff}},\ }\href {\doibase 10.1103/PhysRevLett.100.185701} {\bibfield
  {journal} {\bibinfo  {journal} {Physical Review Letters}\ }\textbf {\bibinfo
  {volume} {100}},\ \bibinfo {pages} {185701} (\bibinfo {year}
  {2008})}\BibitemShut {NoStop}%
\bibitem [{\citenamefont {Scullard}\ and\ \citenamefont {Ziff}(2010)}]{SZ10}%
  \BibitemOpen
  \bibfield  {author} {\bibinfo {author} {\bibfnamefont {C.~R.}\ \bibnamefont
  {Scullard}}\ and\ \bibinfo {author} {\bibfnamefont {R.~M.}\ \bibnamefont
  {Ziff}},\ }\href {http://stacks.iop.org/1742-5468/2010/i=03/a=P03021}
  {\bibfield  {journal} {\bibinfo  {journal} {Journal of Statistical Mechanics:
  Theory and Experiment}\ }\textbf {\bibinfo {volume} {2010}},\ \bibinfo
  {pages} {P03021} (\bibinfo {year} {2010})}\BibitemShut {NoStop}%
\bibitem [{\citenamefont {Ziff}\ \emph {et~al.}(2012)\citenamefont {Ziff},
  \citenamefont {Scullard}, \citenamefont {Wierman},\ and\ \citenamefont
  {Sedlock}}]{ZSWS12}%
  \BibitemOpen
  \bibfield  {author} {\bibinfo {author} {\bibfnamefont {R.~M.}\ \bibnamefont
  {Ziff}}, \bibinfo {author} {\bibfnamefont {C.~R.}\ \bibnamefont {Scullard}},
  \bibinfo {author} {\bibfnamefont {J.~C.}\ \bibnamefont {Wierman}}, \ and\
  \bibinfo {author} {\bibfnamefont {M.~R.~A.}\ \bibnamefont {Sedlock}},\ }\href
  {http://stacks.iop.org/1751-8121/45/i=49/a=494005} {\bibfield  {journal}
  {\bibinfo  {journal} {Journal of Physics A: Mathematical and Theoretical}\
  }\textbf {\bibinfo {volume} {45}},\ \bibinfo {pages} {494005} (\bibinfo
  {year} {2012})}\BibitemShut {NoStop}%
\bibitem [{\citenamefont {Pant}\ \emph {et~al.}(2017)\citenamefont {Pant},
  \citenamefont {Krovi}, \citenamefont {Towsley}, \citenamefont {Tassiulas},
  \citenamefont {Jiang}, \citenamefont {Basu}, \citenamefont {Englund},\ and\
  \citenamefont {Guha}}]{PKT+17}%
  \BibitemOpen
  \bibfield  {author} {\bibinfo {author} {\bibfnamefont {M.}~\bibnamefont
  {Pant}}, \bibinfo {author} {\bibfnamefont {H.}~\bibnamefont {Krovi}},
  \bibinfo {author} {\bibfnamefont {D.}~\bibnamefont {Towsley}}, \bibinfo
  {author} {\bibfnamefont {L.}~\bibnamefont {Tassiulas}}, \bibinfo {author}
  {\bibfnamefont {L.}~\bibnamefont {Jiang}}, \bibinfo {author} {\bibfnamefont
  {P.}~\bibnamefont {Basu}}, \bibinfo {author} {\bibfnamefont {D.}~\bibnamefont
  {Englund}}, \ and\ \bibinfo {author} {\bibfnamefont {S.}~\bibnamefont
  {Guha}},\ }\href {https://arxiv.org/abs/1708.07142} {\bibfield  {journal}
  {\bibinfo  {journal} {arXiv:1708.07142}\ } (\bibinfo {year}
  {2017})}\BibitemShut {NoStop}%
\bibitem [{\citenamefont {Pirandola}(2017)}]{P17}%
  \BibitemOpen
  \bibfield  {author} {\bibinfo {author} {\bibfnamefont {S.}~\bibnamefont
  {Pirandola}},\ }\href {https://arxiv.org/abs/1601.00966v4} {\bibfield
  {journal} {\bibinfo  {journal} {arXiv:1601.00966v4}\ } (\bibinfo {year}
  {2017})}\BibitemShut {NoStop}%
\bibitem [{\citenamefont {Duan}\ \emph {et~al.}(2001)\citenamefont {Duan},
  \citenamefont {Lukin}, \citenamefont {Cirac},\ and\ \citenamefont
  {Zoller}}]{DLCZ01}%
  \BibitemOpen
  \bibfield  {author} {\bibinfo {author} {\bibfnamefont {L.-M.}\ \bibnamefont
  {Duan}}, \bibinfo {author} {\bibfnamefont {M.~D.}\ \bibnamefont {Lukin}},
  \bibinfo {author} {\bibfnamefont {J.~I.}\ \bibnamefont {Cirac}}, \ and\
  \bibinfo {author} {\bibfnamefont {P.}~\bibnamefont {Zoller}},\ }\href
  {http://dx.doi.org/10.1038/35106500} {\bibfield  {journal} {\bibinfo
  {journal} {Nature}\ }\textbf {\bibinfo {volume} {414}} (\bibinfo {year}
  {2001})}\BibitemShut {NoStop}%
\bibitem [{\citenamefont {Duan}\ and\ \citenamefont {Kimble}(2003)}]{DK03}%
  \BibitemOpen
  \bibfield  {author} {\bibinfo {author} {\bibfnamefont {L.-M.}\ \bibnamefont
  {Duan}}\ and\ \bibinfo {author} {\bibfnamefont {H.~J.}\ \bibnamefont
  {Kimble}},\ }\href {\doibase 10.1103/PhysRevLett.90.253601} {\bibfield
  {journal} {\bibinfo  {journal} {Physical Review Letters}\ }\textbf {\bibinfo
  {volume} {90}},\ \bibinfo {pages} {253601} (\bibinfo {year}
  {2003})}\BibitemShut {NoStop}%
\bibitem [{\citenamefont {Barrett}\ and\ \citenamefont {Kok}(2005)}]{BK05}%
  \BibitemOpen
  \bibfield  {author} {\bibinfo {author} {\bibfnamefont {S.~D.}\ \bibnamefont
  {Barrett}}\ and\ \bibinfo {author} {\bibfnamefont {P.}~\bibnamefont {Kok}},\
  }\href {\doibase 10.1103/PhysRevA.71.060310} {\bibfield  {journal} {\bibinfo
  {journal} {Physical Review A}\ }\textbf {\bibinfo {volume} {71}},\ \bibinfo
  {pages} {060310} (\bibinfo {year} {2005})}\BibitemShut {NoStop}%
\bibitem [{\citenamefont {van Dam}\ \emph {et~al.}(2017)\citenamefont {van
  Dam}, \citenamefont {Humphreys}, \citenamefont {Rozpedek}, \citenamefont
  {Wehner},\ and\ \citenamefont {Hanson}}]{DHR17}%
  \BibitemOpen
  \bibfield  {author} {\bibinfo {author} {\bibfnamefont {S.~B.}\ \bibnamefont
  {van Dam}}, \bibinfo {author} {\bibfnamefont {P.~C.}\ \bibnamefont
  {Humphreys}}, \bibinfo {author} {\bibfnamefont {F.}~\bibnamefont {Rozpedek}},
  \bibinfo {author} {\bibfnamefont {S.}~\bibnamefont {Wehner}}, \ and\ \bibinfo
  {author} {\bibfnamefont {R.}~\bibnamefont {Hanson}},\ }\href
  {http://stacks.iop.org/2058-9565/2/i=3/a=034002} {\bibfield  {journal}
  {\bibinfo  {journal} {Quantum Science and Technology}\ }\textbf {\bibinfo
  {volume} {2}},\ \bibinfo {pages} {034002} (\bibinfo {year}
  {2017})}\BibitemShut {NoStop}%
\bibitem [{\citenamefont {Walln\"ofer}\ \emph {et~al.}(2016)\citenamefont
  {Walln\"ofer}, \citenamefont {Zwerger}, \citenamefont {Muschik},
  \citenamefont {Sangouard},\ and\ \citenamefont {D\"ur}}]{WZM+16}%
  \BibitemOpen
  \bibfield  {author} {\bibinfo {author} {\bibfnamefont {J.}~\bibnamefont
  {Walln\"ofer}}, \bibinfo {author} {\bibfnamefont {M.}~\bibnamefont
  {Zwerger}}, \bibinfo {author} {\bibfnamefont {C.}~\bibnamefont {Muschik}},
  \bibinfo {author} {\bibfnamefont {N.}~\bibnamefont {Sangouard}}, \ and\
  \bibinfo {author} {\bibfnamefont {W.}~\bibnamefont {D\"ur}},\ }\href
  {\doibase 10.1103/PhysRevA.94.052307} {\bibfield  {journal} {\bibinfo
  {journal} {Physical Review A}\ }\textbf {\bibinfo {volume} {94}},\ \bibinfo
  {pages} {052307} (\bibinfo {year} {2016})}\BibitemShut {NoStop}%
\bibitem [{\citenamefont {Epping}\ \emph
  {et~al.}(2016{\natexlab{a}})\citenamefont {Epping}, \citenamefont
  {Kampermann},\ and\ \citenamefont {Bru\ss}}]{EKB16}%
  \BibitemOpen
  \bibfield  {author} {\bibinfo {author} {\bibfnamefont {M.}~\bibnamefont
  {Epping}}, \bibinfo {author} {\bibfnamefont {H.}~\bibnamefont {Kampermann}},
  \ and\ \bibinfo {author} {\bibfnamefont {D.}~\bibnamefont {Bru\ss}},\ }\href
  {http://stacks.iop.org/1367-2630/18/i=5/a=053036} {\bibfield  {journal}
  {\bibinfo  {journal} {New Journal of Physics}\ }\textbf {\bibinfo {volume}
  {18}},\ \bibinfo {pages} {053036} (\bibinfo {year}
  {2016}{\natexlab{a}})}\BibitemShut {NoStop}%
\bibitem [{\citenamefont {Epping}\ \emph
  {et~al.}(2016{\natexlab{b}})\citenamefont {Epping}, \citenamefont
  {Kampermann},\ and\ \citenamefont {Bru\ss}}]{EKB16b}%
  \BibitemOpen
  \bibfield  {author} {\bibinfo {author} {\bibfnamefont {M.}~\bibnamefont
  {Epping}}, \bibinfo {author} {\bibfnamefont {H.}~\bibnamefont {Kampermann}},
  \ and\ \bibinfo {author} {\bibfnamefont {D.}~\bibnamefont {Bru\ss}},\ }\href
  {http://stacks.iop.org/1367-2630/18/i=10/a=103052} {\bibfield  {journal}
  {\bibinfo  {journal} {New Journal of Physics}\ }\textbf {\bibinfo {volume}
  {18}},\ \bibinfo {pages} {103052} (\bibinfo {year}
  {2016}{\natexlab{b}})}\BibitemShut {NoStop}%
\bibitem [{\citenamefont {Briegel}\ and\ \citenamefont
  {Raussendorf}(2001)}]{BR01}%
  \BibitemOpen
  \bibfield  {author} {\bibinfo {author} {\bibfnamefont {H.~J.}\ \bibnamefont
  {Briegel}}\ and\ \bibinfo {author} {\bibfnamefont {R.}~\bibnamefont
  {Raussendorf}},\ }\href {\doibase 10.1103/PhysRevLett.86.910} {\bibfield
  {journal} {\bibinfo  {journal} {Physical Review Letters}\ }\textbf {\bibinfo
  {volume} {86}},\ \bibinfo {pages} {910} (\bibinfo {year} {2001})}\BibitemShut
  {NoStop}%
\bibitem [{\citenamefont {Acin}\ \emph {et~al.}(2007)\citenamefont {Acin},
  \citenamefont {Cirac},\ and\ \citenamefont {Lewenstein}}]{ACL07}%
  \BibitemOpen
  \bibfield  {author} {\bibinfo {author} {\bibfnamefont {A.}~\bibnamefont
  {Acin}}, \bibinfo {author} {\bibfnamefont {J.~I.}\ \bibnamefont {Cirac}}, \
  and\ \bibinfo {author} {\bibfnamefont {M.}~\bibnamefont {Lewenstein}},\
  }\href {http://dx.doi.org/10.1038/nphys549} {\bibfield  {journal} {\bibinfo
  {journal} {Nature Physics}\ }\textbf {\bibinfo {volume} {3}} (\bibinfo {year}
  {2007})}\BibitemShut {NoStop}%
\bibitem [{\citenamefont {Perseguers}\ \emph {et~al.}(2008)\citenamefont
  {Perseguers}, \citenamefont {Cirac}, \citenamefont {Ac\'{\i}n}, \citenamefont
  {Lewenstein},\ and\ \citenamefont {Wehr}}]{PCA+08}%
  \BibitemOpen
  \bibfield  {author} {\bibinfo {author} {\bibfnamefont {S.}~\bibnamefont
  {Perseguers}}, \bibinfo {author} {\bibfnamefont {J.~I.}\ \bibnamefont
  {Cirac}}, \bibinfo {author} {\bibfnamefont {A.}~\bibnamefont {Ac\'{\i}n}},
  \bibinfo {author} {\bibfnamefont {M.}~\bibnamefont {Lewenstein}}, \ and\
  \bibinfo {author} {\bibfnamefont {J.}~\bibnamefont {Wehr}},\ }\href {\doibase
  10.1103/PhysRevA.77.022308} {\bibfield  {journal} {\bibinfo  {journal}
  {Physical Review A}\ }\textbf {\bibinfo {volume} {77}},\ \bibinfo {pages}
  {022308} (\bibinfo {year} {2008})}\BibitemShut {NoStop}%
\bibitem [{\citenamefont {Cuquet}\ and\ \citenamefont
  {Calsamiglia}(2009)}]{CC09}%
  \BibitemOpen
  \bibfield  {author} {\bibinfo {author} {\bibfnamefont {M.}~\bibnamefont
  {Cuquet}}\ and\ \bibinfo {author} {\bibfnamefont {J.}~\bibnamefont
  {Calsamiglia}},\ }\href {\doibase 10.1103/PhysRevLett.103.240503} {\bibfield
  {journal} {\bibinfo  {journal} {Physical Review Letters}\ }\textbf {\bibinfo
  {volume} {103}},\ \bibinfo {pages} {240503} (\bibinfo {year}
  {2009})}\BibitemShut {NoStop}%
\bibitem [{\citenamefont {Lapeyre}\ \emph {et~al.}(2009)\citenamefont
  {Lapeyre}, \citenamefont {Wehr},\ and\ \citenamefont {Lewenstein}}]{LWL09}%
  \BibitemOpen
  \bibfield  {author} {\bibinfo {author} {\bibfnamefont {G.~J.}\ \bibnamefont
  {Lapeyre}}, \bibinfo {author} {\bibfnamefont {J.}~\bibnamefont {Wehr}}, \
  and\ \bibinfo {author} {\bibfnamefont {M.}~\bibnamefont {Lewenstein}},\
  }\href {\doibase 10.1103/PhysRevA.79.042324} {\bibfield  {journal} {\bibinfo
  {journal} {Physical Review A}\ }\textbf {\bibinfo {volume} {79}},\ \bibinfo
  {pages} {042324} (\bibinfo {year} {2009})}\BibitemShut {NoStop}%
\end{thebibliography}%

\appendix

\section{Bell measurement calculation}\label{appendix-bell_meas_calc}

	For any $a_3,b_3\in\{0,1\}$, consider the following inner product:
	\begin{align}
		&_{\bar{B}_1\bar{B}_2}\bra{\Phi_{a_3,b_3}}\(\ket{\Phi_{a_1,b_1}}_{B_1\bar{B}_1}\otimes\ket{\Phi_{a_2,b_2}}_{B_2\bar{B}_2}\)\label{eq-ent_swap_pm_state}\\
		&=_{\bar{B}_1\bar{B}_2}\bra{\Phi_{a_3,b_3}}\left((\sigma_x^{a_1}\sigma_z^{b_1}\otimes\mathbbm{1})\ket{\Psi^+}_{B_1\bar{B}_1}\right.\nonumber\\
		&\quad \left.\otimes(\sigma_x^{a_2}\sigma_z^{b_2}\otimes\mathbbm{1})\ket{\Psi^+}_{B_2\bar{B}_2}\right)\nonumber\\
		&=(\sigma_x^{a_1}\sigma_z^{b_1}\otimes\sigma_x^{a_2}\sigma_z^{b_2})~_{\bar{B}_1\bar{B}_2}\bra{\Phi_{a_3,b_3}}\(\ket{\Psi^+}_{B_1\bar{B}_1}\otimes\ket{\Psi^+}_{B_2\bar{B}_2}\)\nonumber\\
		&=(\sigma_x^{a_1}\sigma_z^{b_1}\otimes\sigma_x^{a_2}\sigma_z^{b_2})\frac{1}{2}\ket{\Phi_{a_3,b_3}}_{B_1B_2}\label{eq-ent_swap_pm_state_2}.
	\end{align}
	To obtain the last equality, we used the fact that
	\begin{equation}
		_{\bar{B}_1\bar{B}_2}\bra{\Phi_{a_3,b_3}}\(\ket{\Psi^+}_{B_1\bar{B}_1}\otimes\ket{\Psi^+}_{B_2\bar{B}_2}\)=\frac{1}{2}\ket{\Phi_{a_3,b_3}}_{B_1B_2}
	\end{equation}
	for all $a_3,b_3\in\{0,1\}$ (up to a possible irrelevant global phase), which is the known result from standard entanglement swapping \cite{BBC+93,ZZH93}. We now use the following facts about the Pauli-$x$ and Pauli-$z$ operators to simplify \eqref{eq-ent_swap_pm_state_2}:
	\begin{equation}\label{eq-Pauli_identities}
	\begin{aligned}
		\sigma_z^a\sigma_x^b&=(-1)^{1\oplus a\oplus b}\sigma_x^b\sigma_z^a,\\
		\sigma_x^a\sigma_x^b&=\sigma_x^{a\oplus b},\\
		\sigma_z^a\sigma_z^b&=\sigma_z^{a\oplus b}
	\end{aligned}
	\end{equation}
	for all $a,b\in\{0,1\}$, where $\oplus$ denotes addition modulo two. Therefore, up to possible global phases, we have that
	\begin{align}
		&(\sigma_x^{a_1}\sigma_z^{b_1}\otimes\sigma_x^{a_2}\sigma_z^{b_2})\frac{1}{2}\ket{\Phi_{a_3,b_3}}_{B_1B_2}\\
		&=\frac{1}{2}\(\sigma_x^{a_1\oplus a_3}\sigma_z^{b_1\oplus b_3}\otimes \sigma_x^{a_2}\sigma_z^{b_2}\)\ket{\Psi^+}_{B_1B_2}\\
		&=\frac{1}{2}\(\sigma_x^{a_1\oplus a_3}\sigma_z^{b_1\oplus b_3}\otimes\sigma_x^{a_2}\sigma_z^{b_2}\sigma_x\)\ket{\Phi^+}_{B_1B_2}.
	\end{align}
	To obtain the last equality we used the fact that $\ket{\Psi^+}=(\mathbbm{1}\otimes\sigma_x)\ket{\Phi^+}$. Now, for any square operator $M$, it holds that
	\begin{equation}
		(\mathbbm{1}\otimes M)\ket{\Phi^+}=(M^{\t}\otimes\mathbbm{1})\ket{\Phi^+}
	\end{equation}
	Using this, along with the identities \eqref{eq-Pauli_identities} and $\sigma_x^{\t}=\sigma_x$ and $\sigma_z^{\t}=\sigma_z$, we obtain up to a possible global phase
	\begin{align}
		&\frac{1}{2}\(\sigma_x^{a_1\oplus a_3}\sigma_z^{b_1\oplus b_3}\otimes\sigma_x^{a_2}\sigma_z^{b_2}\sigma_x\)\ket{\Phi^+}_{B_1B_2}\\
		&=\frac{1}{2}\(\sigma_x^{a_1\oplus a_3}\sigma_z^{b_1\oplus b_3}\sigma_x\sigma_z^{b_2}\sigma_x^{a_2}\otimes\mathbbm{1}\)\ket{\Phi^+}_{B_1B_2}\\
		&=\frac{1}{2}\(\sigma_x^{a_1\oplus a_3\oplus a_2}\sigma_z^{b_1\oplus b_3\oplus b_2}\otimes\mathbbm{1}\)\ket{\Psi^+}_{B_1B_2}.
	\end{align}
	Therefore, up to a possible global phase
	\begin{align}
		&_{\bar{B}_1\bar{B}_2}\bra{\Phi_{a_3,b_3}}\(\ket{\Phi_{a_1,b_1}}_{B_1\bar{B}_1}\otimes\ket{\Phi_{a_2,b_2}}_{B_2\bar{B}_2}\)\nonumber\\
		&\qquad =\frac{1}{2}\ket{\Phi_{a_1\oplus a_2\oplus a_3,b_1\oplus b_2\oplus b_3}},
	\end{align}
	which tells us that each outcome of the Bell measurement occurs with probability $\frac{1}{4}$ and that the corresponding post-measurement state is a Bell state.

\end{document}